\documentclass[twocolumn]{aastex63}
\usepackage{verbatim}
\usepackage{color}
\usepackage{CMNDSManual}
\usepackage{CMNDS_Count_z}
\usepackage{CMNDS_PhotoZAGN}
\usepackage{CMNDS_PhotoZGal}
\usepackage{CMNDS_AGN_PAIR_FRAC_ebv_aa}
\usepackage{CMNDS_AGN_PAIR_FRAC_ssfr_aa}
\usepackage{CMNDS_AGN_PAIR_FRAC_ssfr_bb}

\shorttitle{WISE AGN Pairs}
\shortauthors{Barrows et al.}

\begin{document}

\accepted{for publication in ApJ}

\title{A Census of \wise-selected Dual and Offset AGN Across the Sky: New Constraints on Merger-Driven Triggering of Obscured AGN}

\author[0000-0002-6212-7328]{R. Scott Barrows}
\affiliation{Department of Astrophysical and Planetary Sciences, University of Colorado Boulder, Boulder, CO 80309, USA}

\author{Julia M. Comerford}
\affiliation{Department of Astrophysical and Planetary Sciences, University of Colorado Boulder, Boulder, CO 80309, USA}

\author[0000-0003-2686-9241]{Daniel Stern}
\affiliation{Jet Propulsion Laboratory, California Institute of Technology, 4800 Oak Grove Drive, Pasadena, CA 91109, USA}

\author[0000-0002-9508-3667]{Roberto J. Assef}
\affiliation{Instituto de Estudios Astrof\'isicos, Facultad de Ingenier\'ia y Ciencias, Universidad Diego Portales, Av. Ej\'ercito Libertador 441, Santiago, Chile}

\correspondingauthor{R. Scott Barrows}
\email{Robert.Barrows@Colorado.edu}

\begin{abstract}

Pairs of galaxies hosting active galactic nuclei (AGN) are powerful probes of merger-driven supermassive black hole (SMBH) growth as they can resolve individual AGN and trace mergers over a large range of physical separations. To exploit this on a large scale for the first time for both obscured and unobscured AGN, we use photometric redshifts of AGN selected by the \wisetitle~(\wise) to find probabilistic pairs ($<$100\,kpc separations) across the sky, along with a comparison sample of inactive galaxy pairs. Our final sample of integrated pair probabilities yields \NAAz~AGN-AGN pairs (dual AGN) and \NAGz~AGN-galaxy pairs (offset AGN) with uniformly measured AGN and host galaxy physical properties. We find the fraction of galaxy pairs hosting \wise~AGN is dominated by offset AGN and significantly elevated above that of inactive galaxies for large host stellar masses. We show how the AGN merger fraction directly increases with AGN extinction for both offset and dual AGN, with up to $\sim$40\%~of heavily obscured AGN found in galaxy pairs. Elevated AGN merger fractions coincide with increased host specific star formation rates that suggest merger-driven co-evolution of galaxies and SMBHs. Among dual AGN, the most rapid SMBH growth may occur within the less massive galaxy. Relative to stochastic mechanisms, mergers produce an excess of AGN at increasingly smaller separations, especially for obscured AGN (up to a factor of $\sim$5), and augmented by correlated triggering. Finally, this excess is stronger than for lower luminosity optically-selected AGN, regardless of AGN obscuration level.

\end{abstract}

\keywords{Active galactic nuclei (16) --- Galaxy evolution (594) --- Galaxy interactions (600) --- Galaxy mergers (608) --- Galaxy pairs (610) --- Infrared galaxies (790) --- AGN host galaxies (2017)}

\section{Introduction}
\label{sec:intro}

Supermassive black holes (SMBHs; \MBH\,$=$\,$10^{5}$\,$-$\,$10^{9}$\,\MSun) at the centers of galaxies primarily grow through accretion of matter, during which time they appear as active galactic nuclei (AGN). Theory predicts that fueling of the most massive SMBHs can be driven efficiently by interactions and mergers of gas-rich galaxies that deliver gas and dust to galaxy nuclei through gravitational dynamical forces \citep[e.g.][]{Hernquist:1989,Mihos:1996,Hopkins2008,Van_Wassenhove:2012,Blecha:2017}. Additionally, mergers can induce star-formation and randomize stellar orbits in the host galaxies, contributing to SMBH-galaxy co-evolution \citep[e.g.][]{DiMatteo:2005,Springel:2005a,Capelo:2017}. 

However, a comprehensive observational picture of the role that mergers play in AGN triggering has remained elusive. In particular, whether galaxy mergers are necessary for explaining a significant fraction of the observed AGN population, or if their accretion is largely driven by stochastic processes within their host galaxies, is currently unclear. Indeed, a large number of works based on morphologically-selected mergers find that AGN do not preferentially reside in mergers \citep[e.g.][]{Georgakakis:2009,Kocevski:2012,Simmons:2012,Villforth:2014,Mechtley:2015,Villforth:2016,Farrah:2017}.

Obscuration of AGN \citep[which can be significant in galaxy mergers; e.g.][]{Sanders:1988,Satyapal:2017} is often posited as a possible explanation for tenuous AGN-merger connections. For example, while the sensitivity of X-rays to AGN emission and their high energies make that observational regime effective at studying AGN in merging environments \citep[e.g.][]{Comerford:2015,Barrows:2016,Barrows:2017,Ricci:2017,Ricci:2021,DeRosa:2022}, Compton scattering heavily attenuates X-ray photons at large neutral hydrogen column densities. Therefore, many heavily obscured AGN will elude X-ray surveys (\citealp[e.g.][]{Koss:2022}; though this issue would be mitigated by a sensitive hard X-ray observatory, such as the High Energy X-ray Probe mission concept; \citealt{Madsen:2019}), whereas models predict that Compton thick AGN may account for $\sim$10\,$-$\,40\%~of the AGN population \citep[e.g.][]{Gilli:2007,Treister:2009,Ueda:2014}.

The observational hurdle of detecting obscured AGN on a large and uniform scale is overcome using the \wisetitle~\citep[\wise;][]{Wright:2010}. \wise~has surveyed the entire sky in four mid infrared (MIR) bands - \wone~(\lameff\,$=$\,3.4\,\micron), \wtwo~(\lameff\,$=$\,4.6\,\micron), \wthree~(\lameff\,$=$\,12.1\,\micron), and \wfour~(\lameff\,$\approx$\,22\,\micron) - which are much less sensitive than X-rays to obscuration. Furthermore, the \wone\,$-$\,\wtwo~color efficiently distinguishes between the blackbody spectral shape of galaxies and the typical powerlaw spectral shape of AGN at MIR wavelengths \citep[e.g.][]{Ashby:2009,Assef2010,Jarrett:2011,Stern:2012,Mateos:2012,Assef:2013,Secrest:2015,Assef:2018}, enabling the selection of both obscured and unobscured AGN using \wise~photometry. Indeed, several observational works show that galaxy pairs have an elevated number density of \wise-selected AGN relative to isolated galaxies \citep[e.g.][]{Satyapal:2014,Weston:2017,Goulding:2018,Gao:2020}. 

However, simulations strongly suggest that the link between AGN and galaxy mergers has a complex dependence on host galaxy properties. Specifically, though several studies have noted a dependence of the AGN merger fraction on AGN luminosity \citep[e.g.][]{Treister:2012,Glikman:2015,Barrows:2017}, some simulations and models predict those trends are determined by host galaxy stellar mass \citep[e.g.][]{Steinborn:2016,Weigel:2018}, while others suggest they are driven by merger-induced dynamical forces \citep[e.g.][]{McAlpine:2020,Byrne-Mamahit:2022}. Furthermore, the sensitivity of AGN triggering to mergers is expected to increase toward later merger stages \citep[e.g.][]{Van_Wassenhove:2012,Stickley:2014,Capelo:2015}. Mergers are also predicted to enhance AGN extinction \citep[e.g.][]{Blecha:2017}, but the obscuring material may be strongly connected to the galaxy-wide supply of gas and dust. Testing these predictions requires disentangling the intrinsic AGN and host galaxy physical properties for samples of obscured and unobscured AGN. To this end, we recently constructed a catalog of host galaxies, photometric redshifts, and physical properties for \wise-selected AGN \citep{Barrows:2021} based on broadband spectral energy distributions (SEDs).

In this paper, we use the photometric redshift (\zphot) probability distribution functions (PDFs) of that sample and the methodology employed in probabilistic galaxy merger rate estimates to analyze the effects of galaxy pairing and merging on AGN triggering, obscuration, and accretion rates. The broadband SED models (consisting of AGN and galaxy templates) provide uniform estimates of intrinsic AGN and host galaxy properties. Moreover, the sample is selected such that the AGN are spatially associated with distinct, resolved galaxies to avoid source confusion and biases due to morphological selection. This feature permits us to distinguish between pairs with only one AGN (offset AGN) and those with two AGN (dual AGN) to test the impact of mergers on correlated AGN triggering.

While many examples of serendipitously discovered dual AGN exist \citep[e.g.][]{Komossa2003,Guainazzi2005,Hudson2006,Bianchi2008,Piconcelli2010,Koss:2011,Mazzarella:2012,Hou:2022}, few systematically constructed samples exist \citep[e.g.][]{Liu:2011,Koss:2012,Hou:2020}, and even fewer are sensitive to obscured AGN \citep[e.g.][]{Mueller-Sanchez:2015,Weston:2017}, with all relying on spectroscopic redshifts that severely limit the sample sizes. By incorporating photometric detections and probabilistic pairs across the sky, this analysis takes advantage of the vast numbers of \wise~AGN to identify many previously unknown pairs and to construct the largest sample of galaxy mergers with obscured AGN.

This paper is structured as follows: in Section \ref{sec:procedure} we describe our methodology for computing pair probabilities, in Section \ref{sec:mass_completeness} we describe our measurements of redshift-dependent physical properties and merger fractions, in Section \ref{sec:agn_gal_mstar_ssfr} we test whether mergers contribute significantly to the population of MIR-selected AGN, in Section \ref{sec:coevol} we examine how merger-driven AGN triggering impacts AGN accretion rates, AGN obscuration, and host galaxy star formation, in Section \ref{sec:stage} we examine how mergers enhance triggering of obscured and unobscured AGN throughout the merger process, and in Section \ref{sec:conc} we present our conclusions. Throughout we assume a flat cosmology defined by the nine-year Wilkinson Microwave Anisotropy Probe observations \citep{Hinshaw:2013}: $H_{0}$\,$=$\,69.3\,km\,Mpc$^{-1}$\,s$^{-1}$ and $\Omega_{M}$\,$=$\,0.287.

\begin{figure*}[t!]
\includegraphics[width=\textwidth]{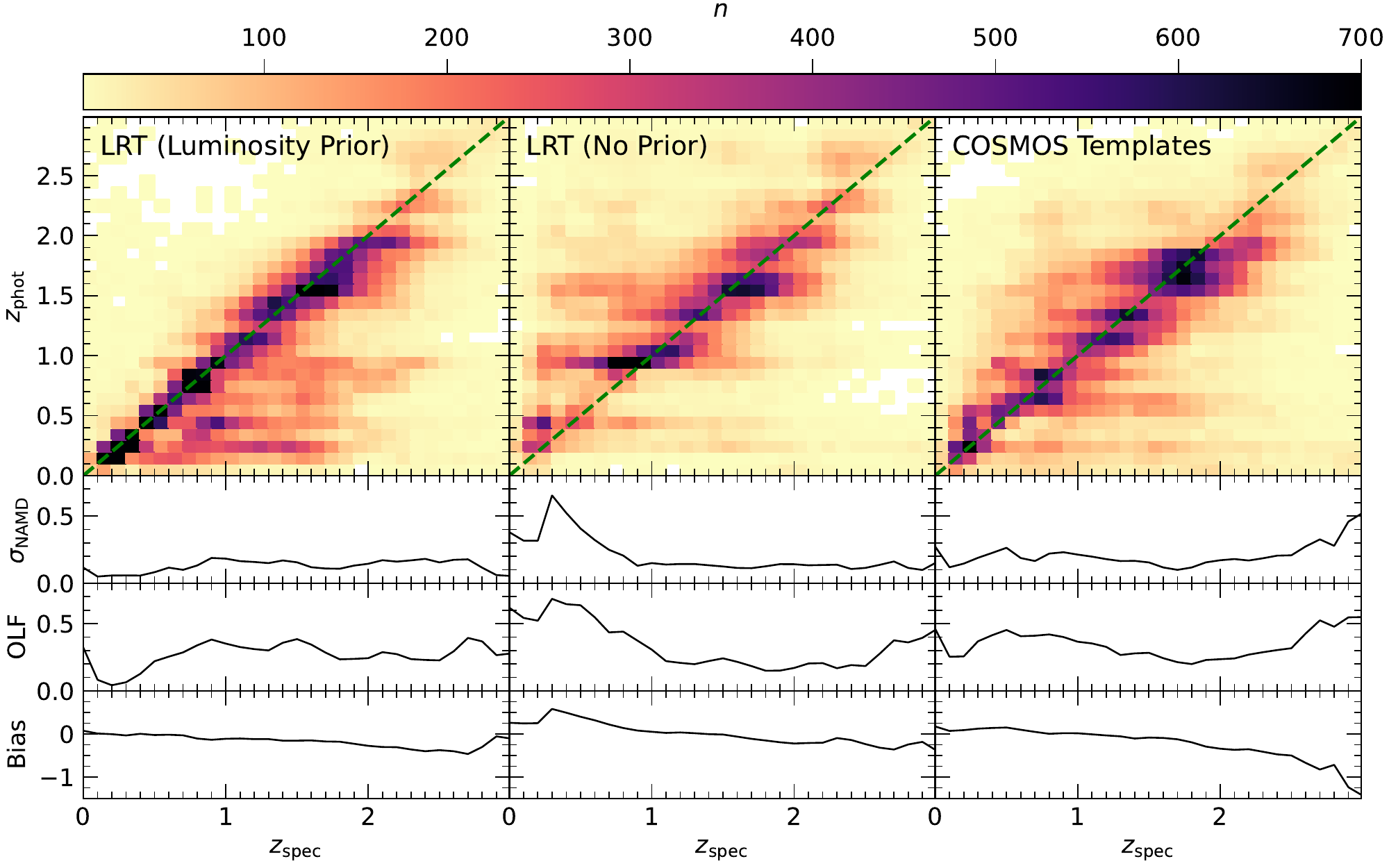}
\caption{\footnotesize{From top to bottom: Photometric redshift (\zphot), standard deviation of the normalized median absolute deviation ($\sigma_{\rm{NMAD}}$\,$=$\,1.48\,$\times$\,median($|\Delta z|/(1+z_{\rm{spec}})$) where $\Delta z=z_{\rm{phot}}-z_{\rm{spec}}$), outlier fraction (OLF; fraction with $|\Delta z|/(1+z_{\rm{spec}})$\,$>$\,0.15), and bias (median $\Delta z$) as a function of spectroscopic redshift (\zspec) for the subset of parent \wise~AGN in the \sdss~DR17 spectroscopic sample. Results are shown from the LRT with and without the luminosity prior applied (left and middle, respectively), and using the COSMOS templates (right). The dashed lines indicate \zphot\,$=$\,\zspec, and the density plot color scale is shown at the top. The overall \zphot~accuracy (based on the $\sigma_{\rm{NMAD}}$ and OLF metrics) is best from the LRT with the luminosity prior applied, but the overall bias is smaller for the other two sets of estimates. See Section \ref{sec:photoz} for details.}}
\label{fig:PHOTOZ_SPECZ}
\end{figure*}

\begin{figure}[t!]
\includegraphics[width=0.48\textwidth]{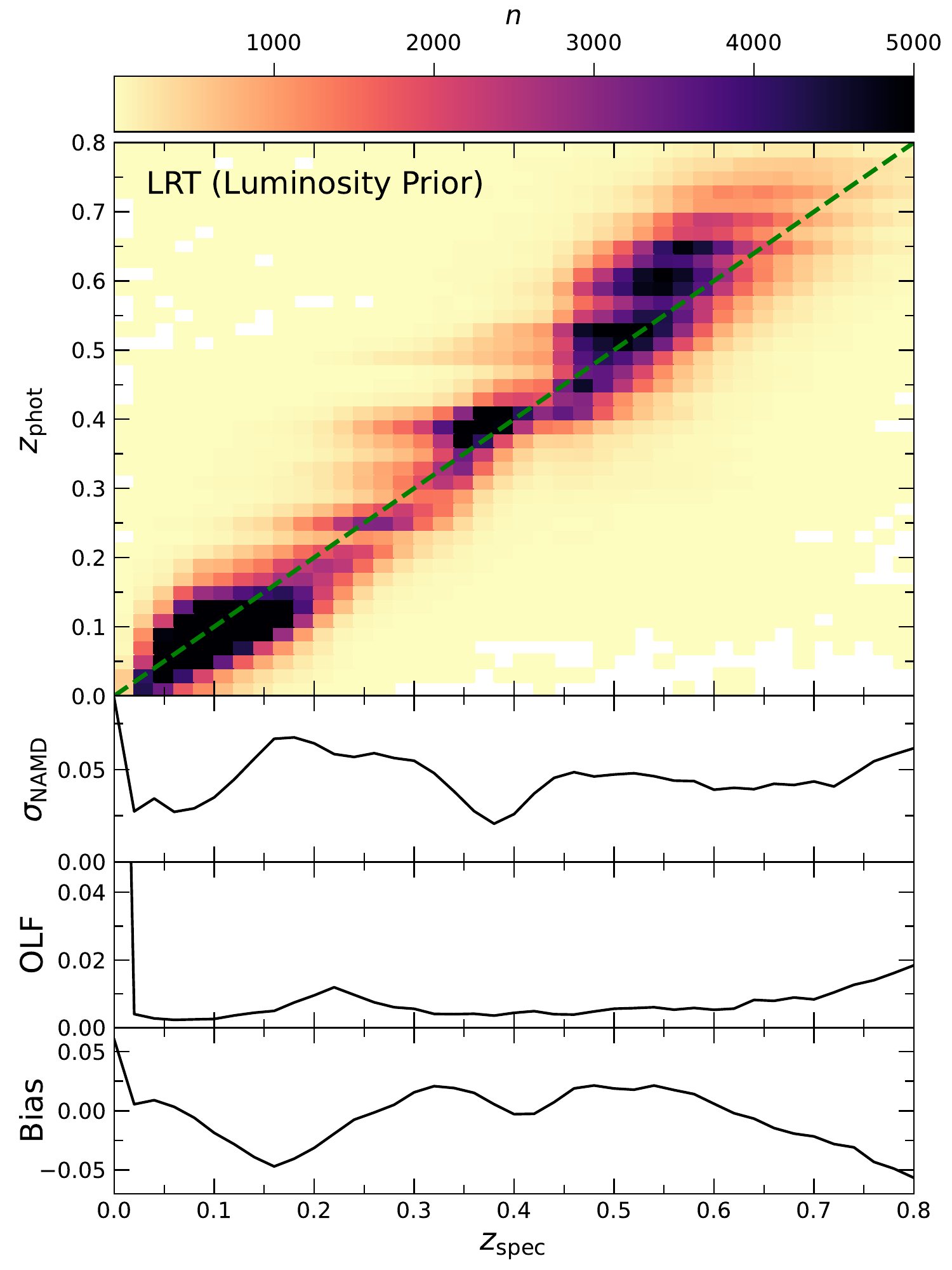}
\caption{\footnotesize{Same as Figure \ref{fig:PHOTOZ_SPECZ} but for the galaxy sample. Results are shown for the template/prior combination adopted for the \wise~AGN (\lrt~with the luminosity prior applied). The dashed line indicates \zphot\,$=$\,\zspec, and the density plot color scale is shown at the top.}}
\label{fig:PHOTOZ_SPECZ_GAL}
\end{figure}

\section{Procedure}
\label{sec:procedure}

Here we describe our procedure. In Section \ref{sec:agn_sample} we provide an overview of the parent sample of \wise~AGN, and in Section \ref{sec:galaxy_sample} we describe the comparison sample of galaxies. In Section \ref{sec:phot_pairs} we detail our methodology for finding photometric pairs. Lastly, in Section \ref{sec:spec} we compare our results with the spectroscopic subset.

\subsection{WISE AGN Sample}
\label{sec:agn_sample}

We use the wide-area catalog of host galaxies for \wise-selected AGN \citep{Barrows:2021} which is based on the AllWISE data release\footnote{https://wise2.ipac.caltech.edu/docs/release/allwise/}. The catalog footprint covers the whole sky above $-$30$^{\circ}$ declination (minus solid angles within 10$^{\circ}$ of the Galactic Plane and 30$^{\circ}$ of the Galactic Center; 23,259\,deg$^2$ total), and consists of 695,273 \wise-selected AGN \citep[90\%~reliability selection from][]{Assef:2018} with photometric redshifts and associated distance properties of the AGN and host galaxies. These estimates were obtained from model fits to broadband SEDs covering ultraviolet to MIR wavelengths. In addition to the \wise~bands, this photometry comes from the \galextitle, the \sdsstitle~(SDSS) Data Release 14 (DR14), \gaia~DR2, the \pnstrstitle~DR2, and the \twomasstitle. We refer the reader to \citet{Barrows:2021} for further details.

\subsection{Galaxy Sample}
\label{sec:galaxy_sample}

To test if AGN are biased toward merging environments relative to inactive galaxies and if mergers have a higher fraction of AGN, we build a sample of galaxies from the SDSS DR17 table of photometric galaxies. We choose this catalog because its footprint provides the largest uniform overlap with the \wise~AGN parent sample and provides similar survey coverage for \zphot~estimates (Section \ref{sec:photoz}). We further limit the galaxies to the same area as the \wise~AGN parent sample (Section \ref{sec:agn_sample}) and require them to satisfy the same \wise~photometric quality criteria applied to the \wise~AGN selection: $W1$\,$>$\,8, $W2$\,$>$\,7, and S/N\,$>$\,5 for $W2$. Doing so builds a galaxy sample that is subject to all selection criteria of the parent \wise~AGN sample except for the MIR AGN color selection (14,975,571 galaxies total).  Within this galaxy sample, 0.7\%~are in the \wise~AGN parent sample, with the remainder not satisfying the \wise~AGN selection criteria.

\subsection{Photometric Pairs}
\label{sec:phot_pairs}

Our procedure for computing pair probabilities uses the full \zphot~PDFs \citep[e.g.][]{Fernandez_Soto:2002,Cunha:2009} to avoid the uncertainties imposed by a strict velocity separation criterion. We follow the methodology of  \citet{Lopez_Sanjuan:2010,Lopez_Sanjuan:2015} and its application to photometric samples for computing galaxy merger fractions \citep[e.g.][]{Duncan:2019,Rodriguez:2020}. This consists of generating \zphot~PDFs (Section \ref{sec:photoz}), correcting the PDFs (Section \ref{sec:photoz_corr}), and computing pair probabilities (Section \ref{sec:pair_prob}).

\subsubsection{Photometric Redshifts}
\label{sec:photoz}

The \zphot~values and PDFs for the parent \wise~AGN sample (Section \ref{sec:agn_sample}) were previously generated in \citet{Barrows:2021} using the \LRTtitle~\citep[\lrt;][]{Assef2010} while applying a physically motivated galaxy luminosity prior \citep{Assef:2008}. Since the pair probability methodology relies critically on the shapes of the PDFs, for comparison we also compute \zphot~values applying two different template/prior combinations to the same sets of photometry. For the first, we use the \lrt~without any priors. For the second, we use a larger set of empirical templates with mixed galaxy and AGN contributions developed from the Cosmic Evolution Survey \citep[COSMOS;][]{Scoville:2007} field \citep{Polletta:2007,Salvato:2009,Salvato:2011}. When using the COSMOS templates, we apply a prior (based on $r-$band magnitude and redshift) that is empirically fit to the data. These additional fits are optimized using LePhare \citep{Arnouts:1999,Ilbert:2006}, and we apply the same reduced chi-squared threshold as in \citet{Barrows:2021}.

The quality for all three sets of \zphot~estimates is quantified graphically in Figure \ref{fig:PHOTOZ_SPECZ} by comparison to spectroscopic redshifts (\zspec; for uniformity, this is limited to the subset within the \sdss~DR17 spectroscopic sample). We quantify the \zphot~accuracy using the normalized median absolute deviation \citep[NMAD;][]{Hoaglin:1983} metric of $\sigma_{\rm{NMAD}}$\,$=$\,1.48\,$\times$\,median($|\Delta z|/(1+z_{\rm{spec}})$), where $\Delta z$\,$=$\,\zphot\,$-$\,\zspec, and the outlier fraction (OLF; defined as the fraction with $|\Delta z|/(1+z_{\rm{spec}})$\,$>$\,0.15). The \zphot~estimates from the \lrt~with the luminosity prior applied ($\sigma_{\rm{NMAD}}$\,$=$\,\SigmaNMADAGNLRTP~and OLF\,$=$\,\OLFAGNLRTP) are overall better than those obtained without the prior ($\sigma_{\rm{NMAD}}$\,$=$\,\SigmaNMADAGNLRTNP~and OLF\,$=$\,\OLFAGNLRTNP) or from the COSMOS templates ($\sigma_{\rm{NMAD}}$\,$=$\,\SigmaNMADAGNCOSMOS~and OLF\,$=$\,\OLFAGNCOSMOS). Hence, we adopt these PDFs when computing pair probabilities. However, as shown in \citet{Assef2010}, while the luminosity prior reduces the overall scatter, it also introduces a systematic offset toward underestimated \zphot~values, resulting in a larger bias (median $\Delta z$\,$=$\,\BiasAGNLRTP) than without the prior (median $\Delta z$\,$=$\,\BiasAGNLRTNP) or with the COSMOS templates (median $\Delta z$\,$=$\,\BiasAGNCOSMOS).

We then compute \zphot~values for the galaxy sample (Section \ref{sec:galaxy_sample}) using the \lrt~with the luminosity prior applied (for uniform comparison with the optimal AGN \zphot~estimates). The galaxy \zphot~accuracy is quantified graphically in Figure \ref{fig:PHOTOZ_SPECZ_GAL} and yields $\sigma_{\rm{NMAD}}$\,$=$\,\SigmaNMADGalLRTP, OLF\,$=$\,\OLFGalLRTP, and bias\,$=$\,\BiasGalLRTP.

\begin{figure}[t!]
\includegraphics[width=0.48\textwidth]{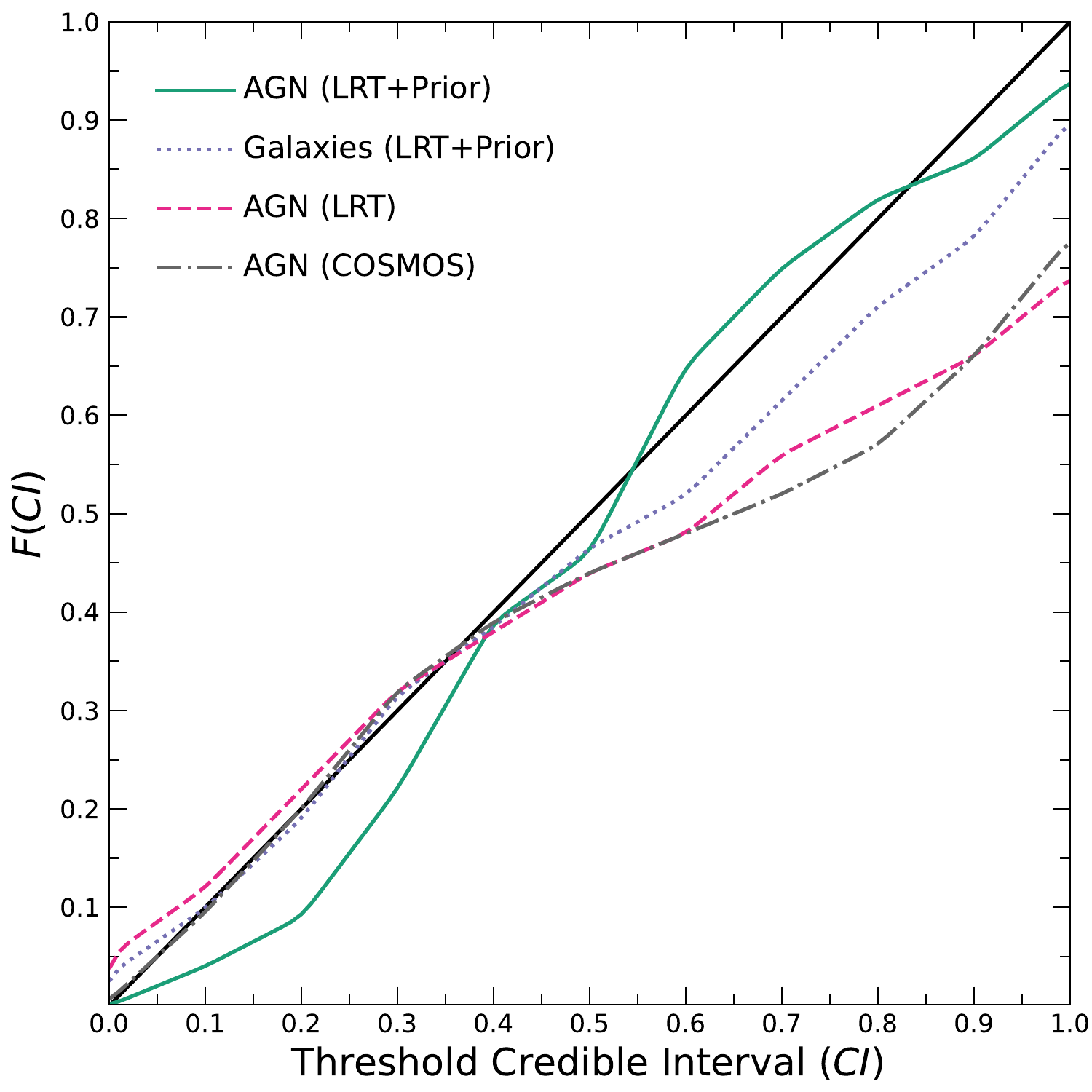} \\
\caption{\footnotesize{Fraction of the spectroscopic subset with spectroscopic redshifts within the threshold credible interval ($F(CI)$) as a function of the threshold credible interval ($CI$; see Section \ref{sec:photoz_corr} for details). $F(CI)$ is plotted for \wise~AGN PDFs from the \lrt~with the luminosity prior applied (green, solid), from the \lrt~without a prior (magenta, dashed), and from the COSMOS templates (gray, dash-dotted). $F(CI)$ is also plotted for the galaxy PDFs from the LRT with the luminosity prior applied (purple, dotted). The solid black line denotes the one-to-one relation expected if the PDFs perfectly reflect the \zphot~uncertainties of the aggregate sample. The best-fit parameters are used to correct the PDFs of the entire \wise~AGN and galaxy samples.}}
\label{fig:QQ}
\end{figure}

\begin{figure}[t!]
\includegraphics[width=0.48\textwidth]{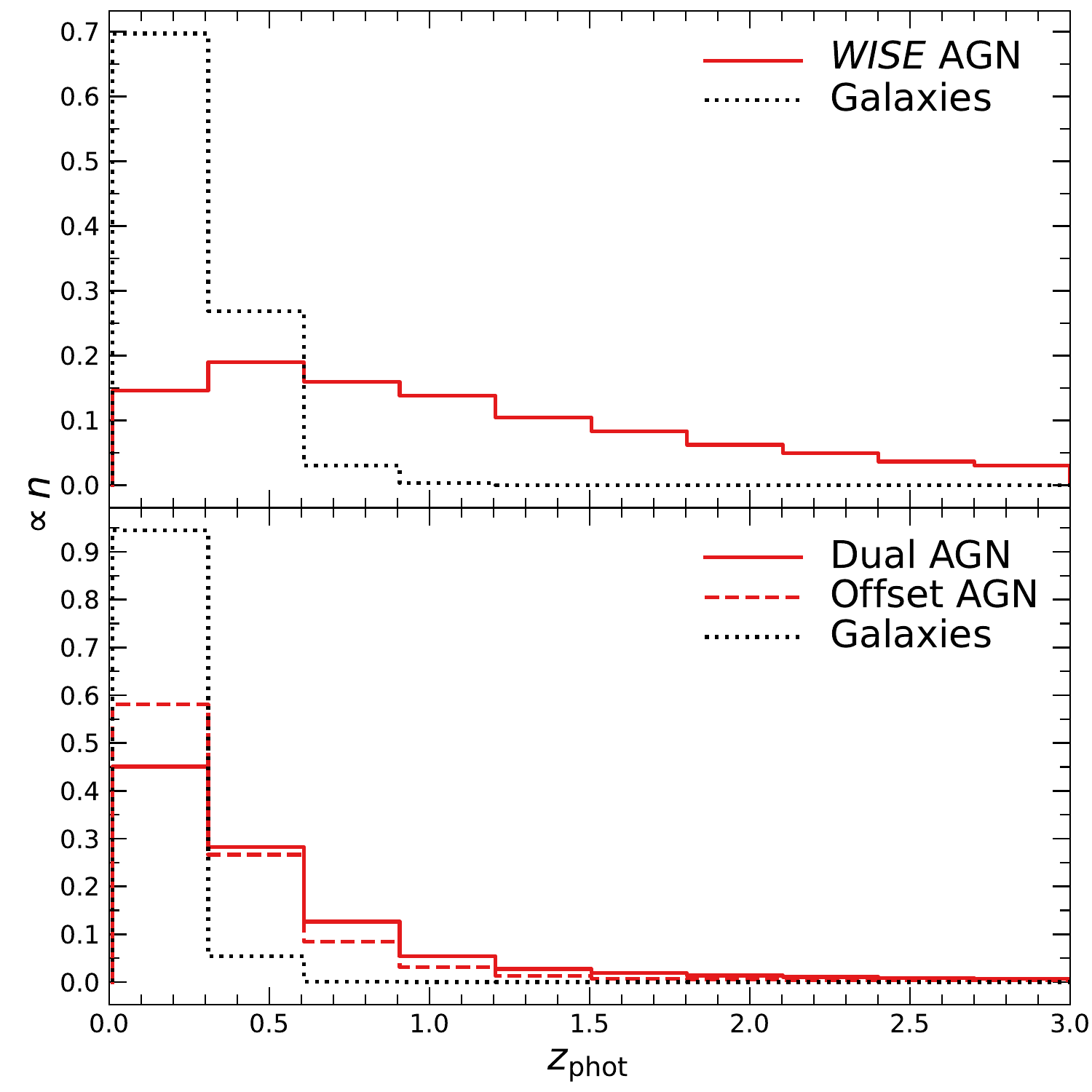}
\caption{\footnotesize{Top: Distribution of photometric redshift (\zphot) probabilities for the \wise~AGN (solid red) and galaxies (dotted black). Bottom: Distribution of pair probabilities for the dual \wise~AGN (solid red), offset \wise~AGN (dashed red), and galaxy pairs (dotted black). For clarity, all subsample distributions are normalized to an integrated value of unity. Relative to the AGN, the galaxies are biased toward low redshifts, and the pair selection biases dual AGN, offset AGN, and galaxy pairs toward lower redshifts.}}
\label{fig:Z}
\end{figure}

\begin{figure}[t!]
\includegraphics[width=0.48\textwidth]{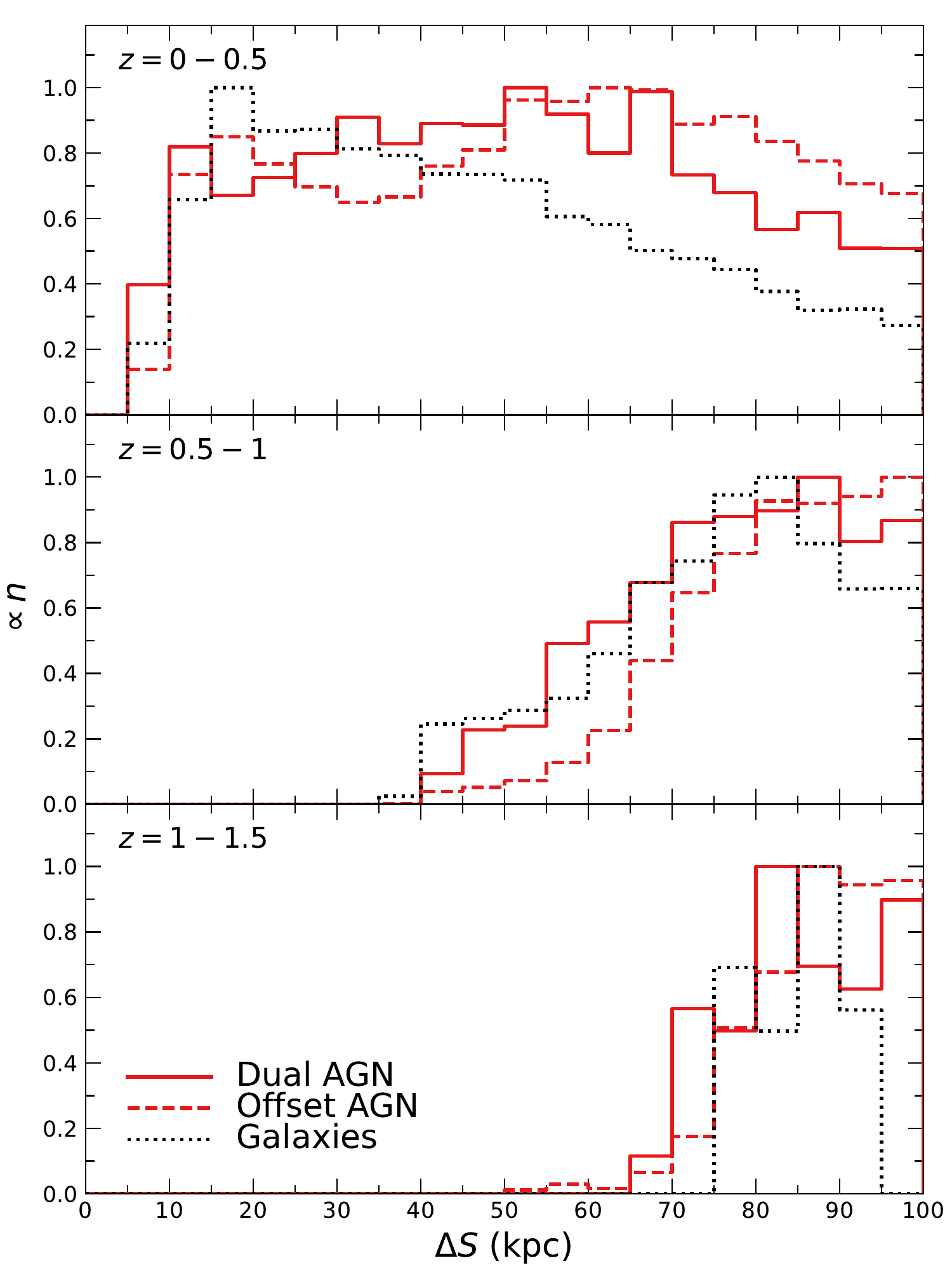}
\caption{\footnotesize{Distribution of probability-weighted projected physical separations (\DeltaS) for pair probabilities in redshift bins of $z$\,$=$\,0\,$-$\,0.5 (top), 0.5\,$-$\,1 (middle), and 1\,$-$\,1.5 (bottom), representing the redshift range that contains $>$\,90\%~of the full pairs sample (Figure \ref{fig:Z}). For clarity, all distributions have been normalized to a maximum of unity. The dual AGN, offset AGN, and galaxy pairs are denoted by the solid red, dashed red, and dotted black lines, respectively. The effect of redshift on the resolvable physical separations is accounted for in Section \ref{sec:mass_completeness}.}}
\label{fig:DELTA_S_MRATIO_with_z}
\end{figure}

\begin{figure}[t!]
\includegraphics[width=0.48\textwidth]{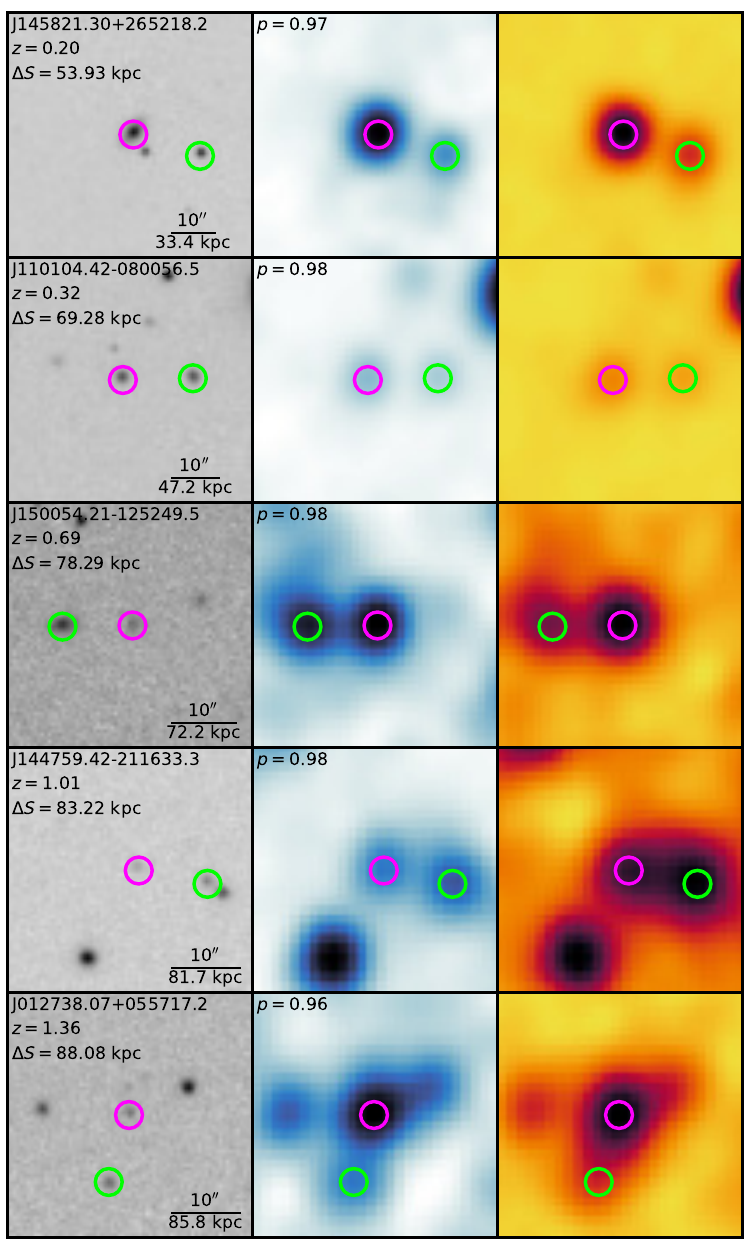}
\caption{\footnotesize{Dual \wise~AGN with pair probabilities $p$\,$>$\,90\%~in five redshift bins defined by the limits $z=[0, 0.3, 0.6, 0.9, 1.2, 1.5]$ (representing the redshift range containing $>$\,90\%~of the full pairs sample; Figure \ref{fig:Z}). The left panel shows the PanSTARRS $i-$band image (highest sensitivity PanSTARRS filter), and the middle and right panels show the \wise~$W1$ and $W2$ images, respectively. For each pair, the adopted redshifts ($z$; labeled) correspond to the maximum value of the pair probability distribution. The examples shown represent the pair with the median projected physical separation (\DeltaS; labeled) in each redshift bin. The magenta and green circles denote the \wise~centroids of the AGN with the larger and smaller bolometric luminosities (\LBol), respectively. Each image is centered on the more luminous AGN (labeled coordinates). The values of \DeltaS~and \LBol~used are based on the adopted, labeled value of $z$.}}
\label{fig:Pairs_AA}
\end{figure}

\begin{figure}[t!]
\includegraphics[width=0.48\textwidth]{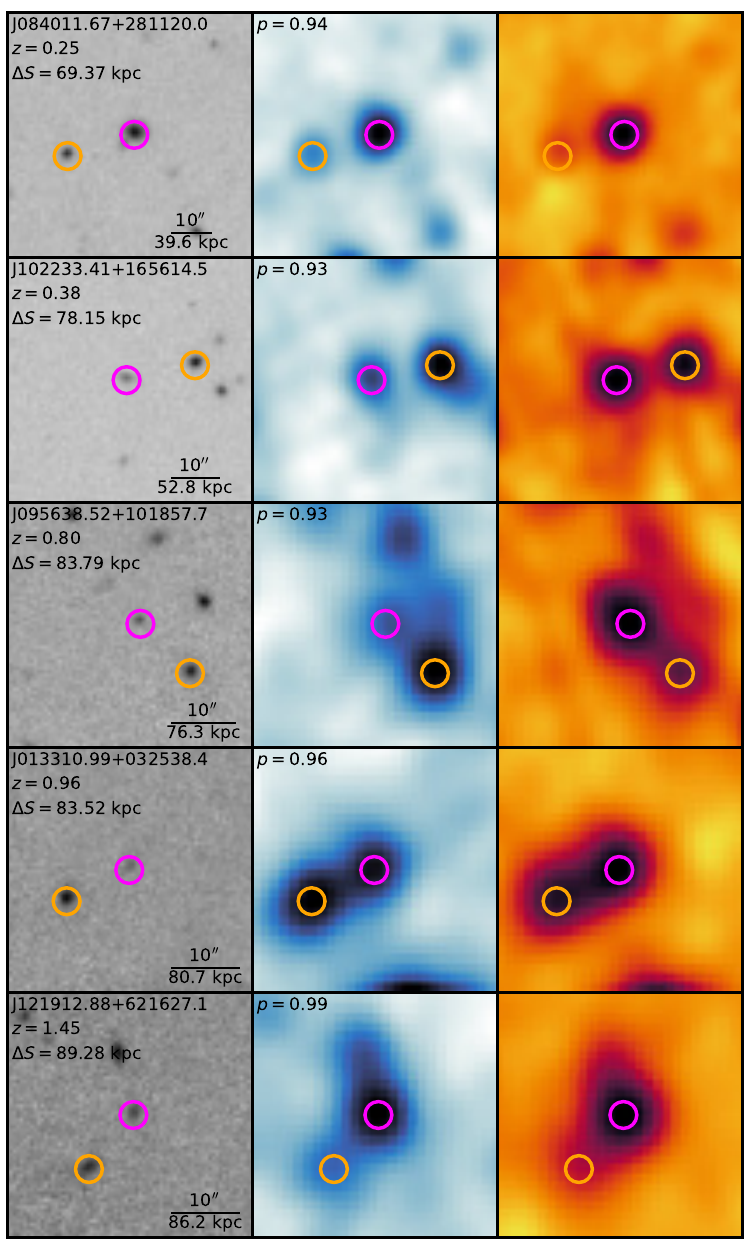}
\caption{\footnotesize{Same as Figure \ref{fig:Pairs_AA} but for offset \wise~AGN. The magenta and orange circles denote the \wise~centroid of the AGN and the companion galaxy $r-$band detection, respectively. Each image is centered on the AGN (labeled coordinates).}}
\label{fig:Pairs_SA}
\end{figure}

\subsubsection{Photometric Redshift PDF Corrections}
\label{sec:photoz_corr}

Due to the inherent variety of SED shapes compared to the templates used to fit them, the \zphot~PDFs will not perfectly reflect the true uncertainties of the \zphot~estimates. These deviations will also be affected by the specific minimization algorithms used (i.e. \lrt~versus LePhare). Therefore, we correct the \zphot~PDFs of the full \wise~AGN and galaxy samples following the procedure from \citet{Wittman:2016} and as implemented in \citet{Duncan:2018}. Using the subset with \zspec~values from the \sdss~DR17, for each PDF (P) we solve for a magnitude-dependent correction (the SDSS spectroscopic coverage is $\sim$25\%~and $\sim$30\%~for the \wise~AGN and galaxy samples, respectively). The corrected PDF is $P_{c}$\,$=$\,$P^{\alpha}$, and $\alpha=\alpha_{c}+\kappa\times(m-m_{c})$ for $m$ greater than the reference magnitude $m_{c}$ (otherwise $\alpha=\alpha_{c}$).  For the reference magnitude, we use the $r-$band and a value of $m_{c}$\,$=$\,18. The magnitude-dependent correction accounts for potential biases in magnitude between the spectroscopic and photometric samples.

We then compute the fraction of sources with \zspec~values within a given credible interval, $CI$: $F(CI)$. For example, the 10\%~$CI$ is the redshift range around the \zphot~PDF peak value within which the integrated probability is equal to 10\%~of the total PDF probability. We then use \texttt{EMCEE} \citep{Foreman-Mackey:2013} to solve for the value of $\alpha$ that minimizes the offset of $F(CI)$ from the expected fraction defined by the $CI$ if the PDFs perfectly represent the \zphot~uncertainties for the aggregate sample (i.e. the one-to-one relation). The results are shown for the \wise~AGN and galaxies in Figure \ref{fig:QQ}.

\subsubsection{Pair Probabilities}
\label{sec:pair_prob}

Following \citet{Lopez_Sanjuan:2015}, we compute the redshift-dependent line-of-sight pair probabilities $Z(z)=\frac{2\times P_{1}(z)\times P_{2}(z)}{P_{1}(z)+P_{2}(z)}$ where the subscripts `1' and `2' denote the two sources in each pair. A source can be part of more than one pair, and each unique pair probability distribution is determined separately.

We compute these probability distributions for pairs of \wise~AGN (dual AGN) and \wise~AGN (within the galaxy sample) paired with galaxies without \wise~AGN (offset AGN). We also compute probabilities for pairs of galaxies without \wise~AGN (galaxy pairs). The total numbers of dual AGN, offset AGN, and galaxy pairs within a given redshift range (\Ndagn, \Noagn, and \Ngg, respectively) are computed by summing the pair probabilities of all unique dual AGN, offset AGN, and galaxy pairs (\ndagn, \noagn, and \ngg, respectively), integrated over the redshift limits ($z_{\rm{min}}$ and $z_{\rm{max}}$): 

\begin{displaymath}
N_{\rm{dAGN}}=\sum_{i=0}^{n_{\rm{dAGN}}}\int_{z=z_{\rm{min}}}^{z_{\rm{max}}}Z_i(z)
\end{displaymath}

\begin{displaymath}
N_{\rm{oAGN}}=\sum_{i=0}^{n_{\rm{oAGN}}}\int_{z=z_{\rm{min}}}^{z_{\rm{max}}}Z_i(z)
\end{displaymath}

\begin{displaymath}
N_{\rm{Gal-Gal}}=\sum_{i=0}^{n_{\rm{Gal-Gal}}}\int_{z=z_{\rm{min}}}^{z_{\rm{max}}}Z_i(z)
\end{displaymath}

For all pair probability functions, the range defined by $z_{\rm{min}}$ and $z_{\rm{max}}$ is restricted to values for which the pair angular separation corresponds to a projected physical separation of \DeltaS\,$\le$\,100\,kpc.  This limit is chosen to investigate the full range of physical separations over which simulations predict that mergers affect galaxy and AGN properties. These criteria yield a total of \NAAz~dual \wise~AGN (\NAAGalAllz~of which are in the \sdss~galaxy sample) and \NAGz~offset \wise~AGN out to $z$\,$=$\,3. Since this study is the first to use these \wise~AGN \zphot~PDFs for pair identification, these dual and offset AGN pair probabilities were previously unknown. For comparison, using a strict velocity offset upper limit of 500 km s$^{-1}$ among the subset with SDSS DR17 spectra, the numbers of spectroscopic dual and offset \wise~AGN are only 6 and 16, respectively.

The total numbers of parent \wise~AGN and galaxies (\Nagn~and \Ngal, respectively) are similarly computed by summing the \zphot~PDFs of all AGN and galaxies (\nagn~and \ngal, respectively), integrated over the same redshift limits:

\begin{displaymath}
N_{\rm{AGN}}=\sum_{i=0}^{n_{\rm{AGN}}}\int_{z=z_{\rm{min}}}^{z_{\rm{max}}}Z_i(z)
\end{displaymath}

\begin{displaymath}
N_{\rm{Gal}}=\sum_{i=0}^{n_{\rm{Gal}}}\int_{z=z_{\rm{min}}}^{z_{\rm{max}}}Z_i(z)
\end{displaymath}

The \zphot~PDFs of the parent \wise~AGN and galaxy samples are shown in Figure \ref{fig:Z}, along with the pair probabilities of the dual AGN, offset AGN, and galaxy pairs. Compared to the AGN, the galaxies exhibit a strong bias toward lower redshifts, and the \DeltaS\,$\le$\,100\,kpc requirement biases all pairs toward lower redshifts. The probability-weighted \DeltaS~values are shown in Figure \ref{fig:DELTA_S_MRATIO_with_z}, where the resolvable physical separations are seen to significantly increase with redshift (this is accounted for when computing merger fractions; Section \ref{sec:mass_completeness}). Figures \ref{fig:Pairs_AA} and \ref{fig:Pairs_SA} show examples of \wise~dual and offset AGN, respectively, with integrated pair probabilities of $p$\,$>$\,0.90.

\begin{figure}[t!]
\includegraphics[width=0.48\textwidth]{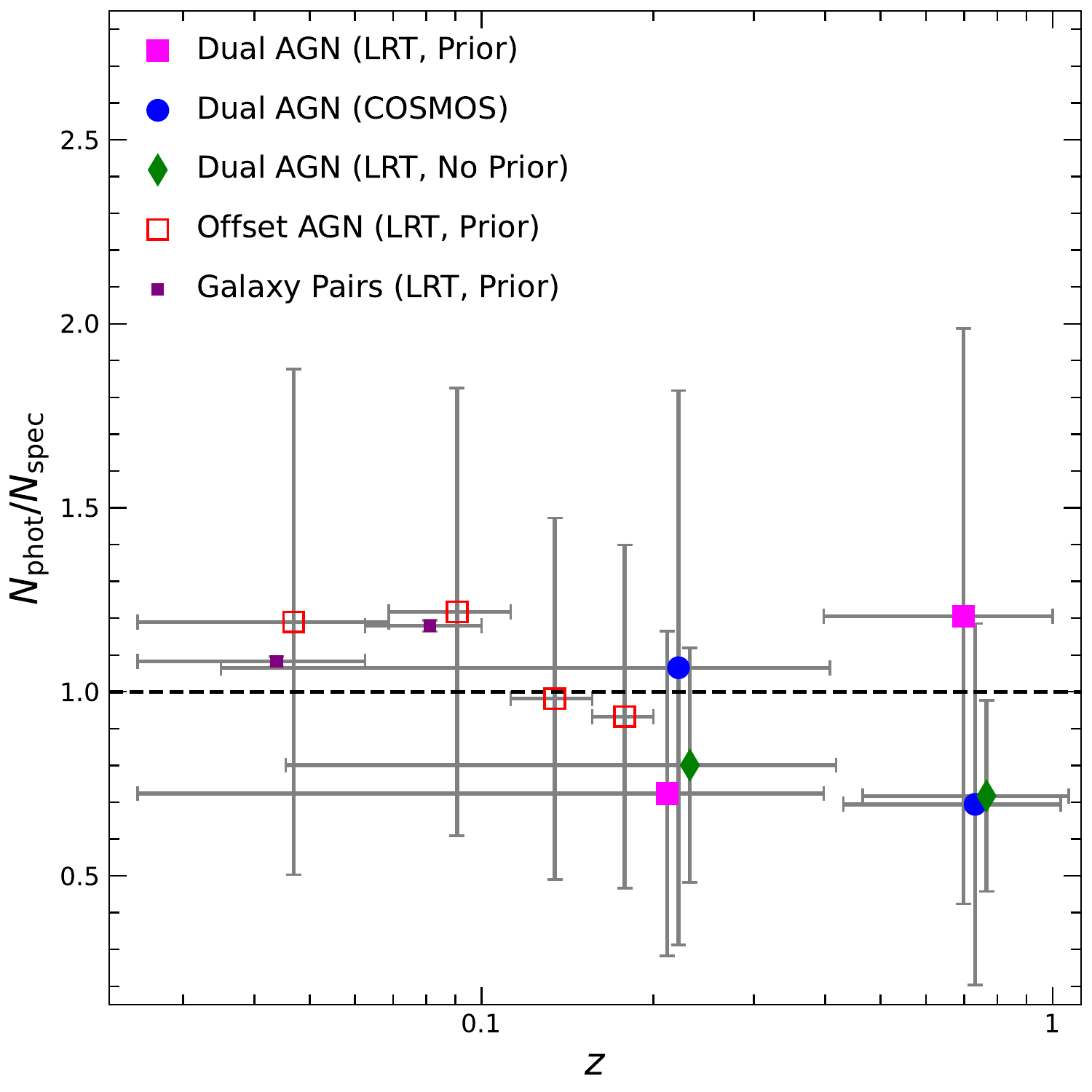}
\caption{\footnotesize{Ratio of the number of photometric pairs ($N_{\rm{phot}}$) to spectroscopic pairs ($N_{\rm{spec}}$; velocity offsets of $<$\,1000\,km\,s$^{-1}$) for projected physical separations of \DeltaS\,$\le$\,100\,kpc. Results from the \lrt~with the luminosity prior applied are denoted by squares (filled magenta, open red, and filled purple for the dual AGN, offset AGN, and galaxy pairs, respectively), while those without the luminosity prior and those using the COSMOS templates are denoted by green diamonds and blue circles, respectively (offset horizontally for clarity). Vertical errorbars denote the Poisson uncertainties, while horizontal errorbars denote the bin widths. The velocity offset threshold and bin sizes are chosen to have at least two spectroscopic pairs per bin.}}
\label{fig:COMPARE_NMERGER}
\end{figure}

\subsection{Comparison with Spectroscopic Pairs}
\label{sec:spec}

To test the efficacy of the procedure described in Section \ref{sec:phot_pairs} for finding photometric pairs, as in \citet{Duncan:2019} we compare the number of pairs identified from photometry against the number based on velocity offsets from the SDSS DR17 \zspec~values. In addition to the \DeltaS\,$\le$\,100\,kpc criterion, we consider spectroscopic pairs to be those with velocity offsets within 1000 km s$^{-1}$ (while this threshold is larger than the commonly adopted value of 500 km s$^{-1}$, we choose it so that at least two spectroscopic sources are in each redshift bin). Photometric pairs are computed as described in Section \ref{sec:pair_prob} from the spectroscopic sample.

The ratio of the number of photometric pairs ($N_{\rm{phot}}$) to the number of spectroscopic pairs ($N_{\rm{spec}}$) is shown in Figure \ref{fig:COMPARE_NMERGER} for dual AGN, offset AGN, and galaxy pairs. Due to the relatively small numbers of spectroscopic dual \wise~AGN, we also show these results using the two additional \zphot~estimates described in Section \ref{sec:photoz}. Relative to the dual AGN, the offset AGN and galaxy pairs probe a lower redshift range (a result of the shallower \sdss~photometric galaxy sample; Figure \ref{fig:Z}). The $N_{\rm{phot}}/N_{\rm{spec}}$ ratios are consistent with unity over the redshift range examined ($z$\,$=$\,0\,$-$\,1, based on the available spectroscopic pairs) for dual AGN, offset AGN, and galaxy pairs, and among the different sets of photometric redshift estimates. For comparison, when adopting a velocity offset threshold of 500 km s$^{-1}$, the ratios are still all consistent with unity, though with larger uncertainties.

\begin{figure}[t!]
\includegraphics[width=0.48\textwidth]{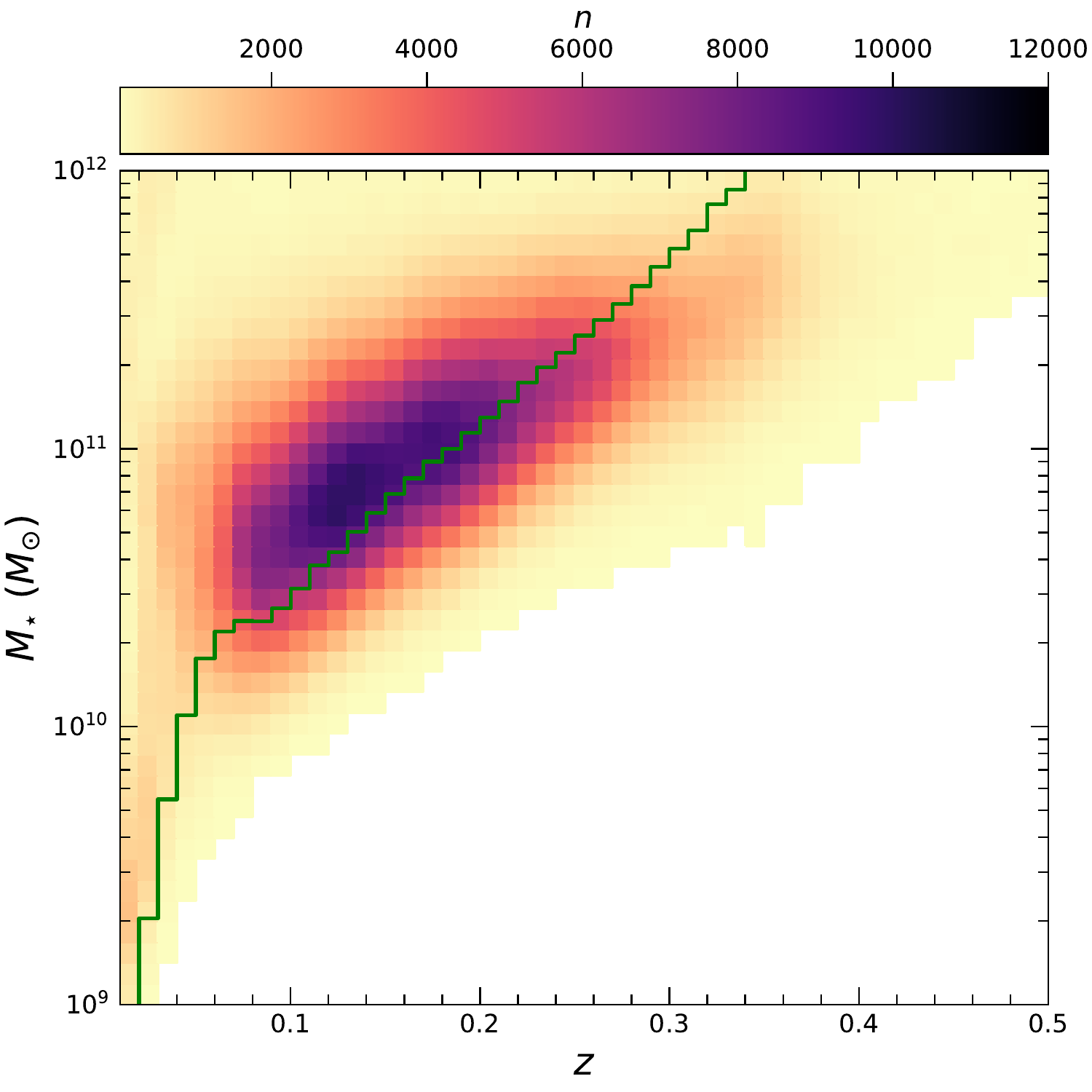}
\caption{\footnotesize{Density plot of the best-fit host galaxy stellar mass (\Mstar) versus photometric redshift (\zphot) for the galaxy sample. We restrict this analysis to $z$\,$\le$\,0.5 (this range encompasses the vast majority of the galaxy pairs; Figure \ref{fig:Z}). The green    line represents the 95th percentile of the lowest mass detectable given the adopted flux limit of $r$\,$=$\,19 (Section \ref{sec:mass_completeness}). The color scale is shown at the top.}}
\label{fig:MCOMP_Z}
\end{figure}

\begin{figure*}[t!]
\includegraphics[width=0.96\textwidth]{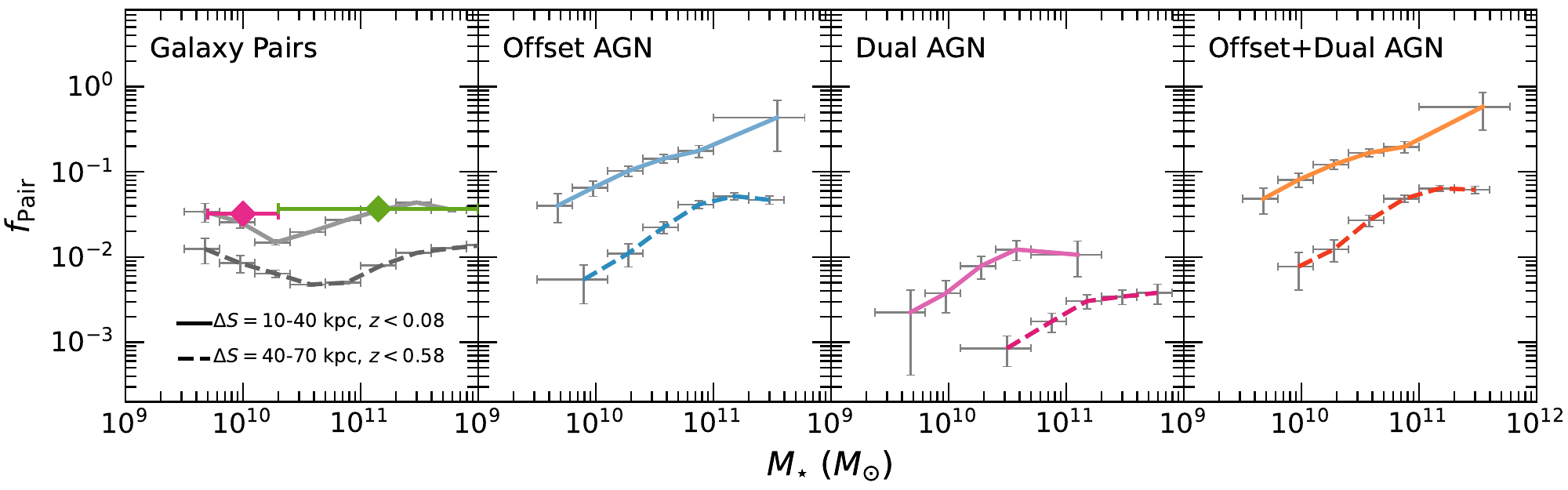}
\caption{\footnotesize{From left to right: Fraction of unique galaxies in pairs (\fgalpair) and fraction of unique \wise~AGN in offset systems, dual systems, and offset or dual systems (\fagno, \fagnd, and \fagnod, respectively) as a function of host galaxy stellar mass (\Mstar). These results are shown for two subsamples in bins of \DeltaS\,$=$\,10\,$-$\,40\,kpc (solid lines) and \DeltaS\,$=$\,40\,$-$\,70\,kpc (dashed lines), limited to $z$\,$\le$\,0.08 and $z$\,$\le$\,0.58 to uniformly resolve down to 10\,kpc and 40\,kpc, respectively, using the 6$''$ resolution of the \wise~source identifications.  All AGN pair fractions (\fagno, \fagnd, and \fagnod) increase with increasing \Mstar. For comparison, the magenta and green diamonds in the left panel denote the galaxy pair fractions from the powerlaw function of \citet{Duncan:2019} for their low-mass and high-mass samples (horizontal errorbars denote their mass ranges).}}
\label{fig:AGN_GAL_PAIR_FRAC}
\end{figure*}

\section{Redshift-Dependent Physical Properties and Mass-Completeness}
\label{sec:mass_completeness}

Since physical distance properties are dependent on the \zphot~estimates, the AGN properties of bolometric luminosity (\LBol) and obscuration (color excess; \EBV) as a function of redshift are determined by fitting the \lrt~models (galaxy templates plus an extinguished AGN template) to the SEDs of the \wise~AGN over a grid of redshifts that covers the full range explored in the \zphot~estimate step (Section \ref{sec:photoz}). Likewise, the host galaxy properties of stellar mass (\Mstar) and star formation rate (SFR) are determined by fitting galaxy models using the Code Investigating GALaxy Emission \citep[\texttt{CIGALE};][]{Noll:2009,Boquien:2019}, as in \citet{Barrows:2021}, to the SEDs of the galaxies and AGN (after subtracting the fitted observed AGN component) over the same redshift grid.

For estimating merger fractions, we define a flux-based and redshift-dependent stellar mass lower limit. Following the procedures outlined in \citet{Pozzetti:2010} and \citet{Darvish:2015}, at each redshift step we compute this limit as the 95th percentile of the limiting masses (Figure \ref{fig:MCOMP_Z}). Since we limit the AGN merger fractions to the subset in the galaxy sample (for comparison with the non-AGN merger fractions), we compute this limit for the galaxy sample and adopt a limit of $r$\,$=$\,19 (this omits the faintest $\sim$10\%~of the galaxies) for computing the limiting masses. We also impose strict lower and upper stellar mass limits of \Mstar\,$=$\,$10^{9}$\,\MSun~and \Mstar\,$=$\,$10^{12}$\,\MSun, respectively. We then compute the fraction of unique AGN in offset and dual systems (\fagno~and \fagnd, respectively) and the fraction of unique galaxies in galaxy pairs (\fgalpair). All sources are subject to the redshift integration limits for which the corresponding values of \Mstar~satisfy these requirements so that they are complete down to a common limiting stellar mass at each redshift. Since these fractions are ratios of individual sources, we can test them as a function of individual galaxy or AGN properties (Sections \ref{sec:agn_gal_mstar_ssfr} and \ref{sec:coevol}).

Furthermore, since the pair samples span a large range of the merger sequence, we develop two subsamples using bins of even size along the \DeltaS~distribution: \DeltaS\,$=$\,10\,$-$\,40\,kpc and \DeltaS\,$=$\,40\,$-$\,70\,kpc. These \DeltaS~bins are chosen to optimize the subsample sizes while also exploring out to separations over which mergers are predicted to influence galaxy properties \citep[e.g.][]{Van_Wassenhove:2012,Stickley:2014,Capelo:2015}. So that the merger fractions are complete down to the \DeltaS~lower limit, we restrict all \zphot~probability distributions of the two subsamples to $z$\,$<$\,0.08 and $z$\,$<$\,0.58, corresponding to \DeltaS\,$=$\,10\,kpc and \DeltaS\,$=$\,40\,kpc for an angular separation of 6$''$ (the resolution limit of \wise~AGN pairs imposed by the $W2$ band detections).

\section{\wise-Selected AGN Are Preferentially Found in Offset or Dual Systems}
\label{sec:agn_gal_mstar_ssfr}

Numerical simulations ubiquitously demonstrate that the dynamics of galaxy mergers can efficiently trigger AGN \citep[e.g.][]{Hernquist:1989,Mihos:1996,Hopkins2008,Van_Wassenhove:2012,Capelo:2015,Blecha:2017}. However, AGN can also be triggered by stochastic processes that occur within galaxies \citep[e.g.][]{Lynden-Bell:1979,Sellwood:1981,VanAldaba:1981,Combes:1985,Pfenniger:1991,Heller:1994,Bournaud:2002,Athanassoula:2003,Sakamoto:1999}. If these internal galaxy processes are the dominant mechanisms for triggering accretion onto SMBHs, then galaxy mergers may be largely incidental to the AGN population.

Investigating the role of galaxy mergers among AGN requires comparison with control samples of galaxies without AGN to determine if accreting SMBHs show a preference for existing in merging environments. Variations of this test have been used in numerous studies and with a wide range of results. Some find an elevated fraction of AGN in mergers compared to non-mergers \citep[e.g.][]{Ellison:2011,Silverman:2011,Ellison:2013,Satyapal:2014,Weston:2017,Goulding:2018,Gao:2020}, while others find no AGN fraction enhancement for galaxies with disturbed morphologies \citep[e.g.][]{Kocevski:2012,Villforth:2014,Mechtley:2015,Villforth:2016,Jin:2021,Silva:2021} or in disk-dominated galaxies \citep[e.g.][]{Georgakakis:2009,Simmons:2012}. These disagreements span a variety of morphological selection methods and exist for AGN selected from both optical and X-ray observations. However, as discussed in Section \ref{sec:intro}, studies utilizing MIR AGN selection do generally find a preference for AGN to be found in mergers \citep[e.g.][]{Satyapal:2014,Weston:2017,Goulding:2018,Gao:2020} due to the inclusion of obscured AGN. Still, AGN fractions may depend strongly on host stellar mass \citep[e.g.][]{Kartaltep:2010,Aird:2012}, and testing this dependence requires disentangling the relative contributions of the AGN and host galaxies among mergers. Our sample, with uniform SED models that account for the presence of AGN, is ideal for this purpose.

Figure \ref{fig:AGN_GAL_PAIR_FRAC} shows \fgalpair, \fagno, \fagnd, and \fagnod~(the fraction of AGN in either offset or dual systems) as a function of \Mstar. The left panel shows that \fgalpair~is broadly consistent with current estimates of the galaxy merger or pair fractions as estimated from a variety of methods in the literature \citep[e.g.][]{Qu:2017,Mundy:2017,Duncan:2019}. Above \Mstar\,$\sim$\,$10^{10}$\,\MSun, a potential positive trend is seen for \fgalpair~with \Mstar. Indeed, some works do suggest such a positive dependence for major mergers \citep[e.g.][]{Weigel:2017,Wang:2020,Nevin:2023}. However, the next three panels show that \fagno, \fagnd, and \fagnod~have a stronger positive dependence with \Mstar. Additionally, \fagnd~is approximately one order of magnitude lower than \fagno~across the range of stellar masses examined (the implications for correlated AGN triggering are discussed further in Section \ref{sec:stage}). These results are qualitatively similar for both the \DeltaS\,$=$\,10\,$-$\,40\,kpc and \DeltaS\,$=$\,40\,$-$\,70\,kpc subsamples, though the merger fractions are overall higher for the smaller separation pairs.

\begin{figure}[t!]
\includegraphics[width=0.48\textwidth]{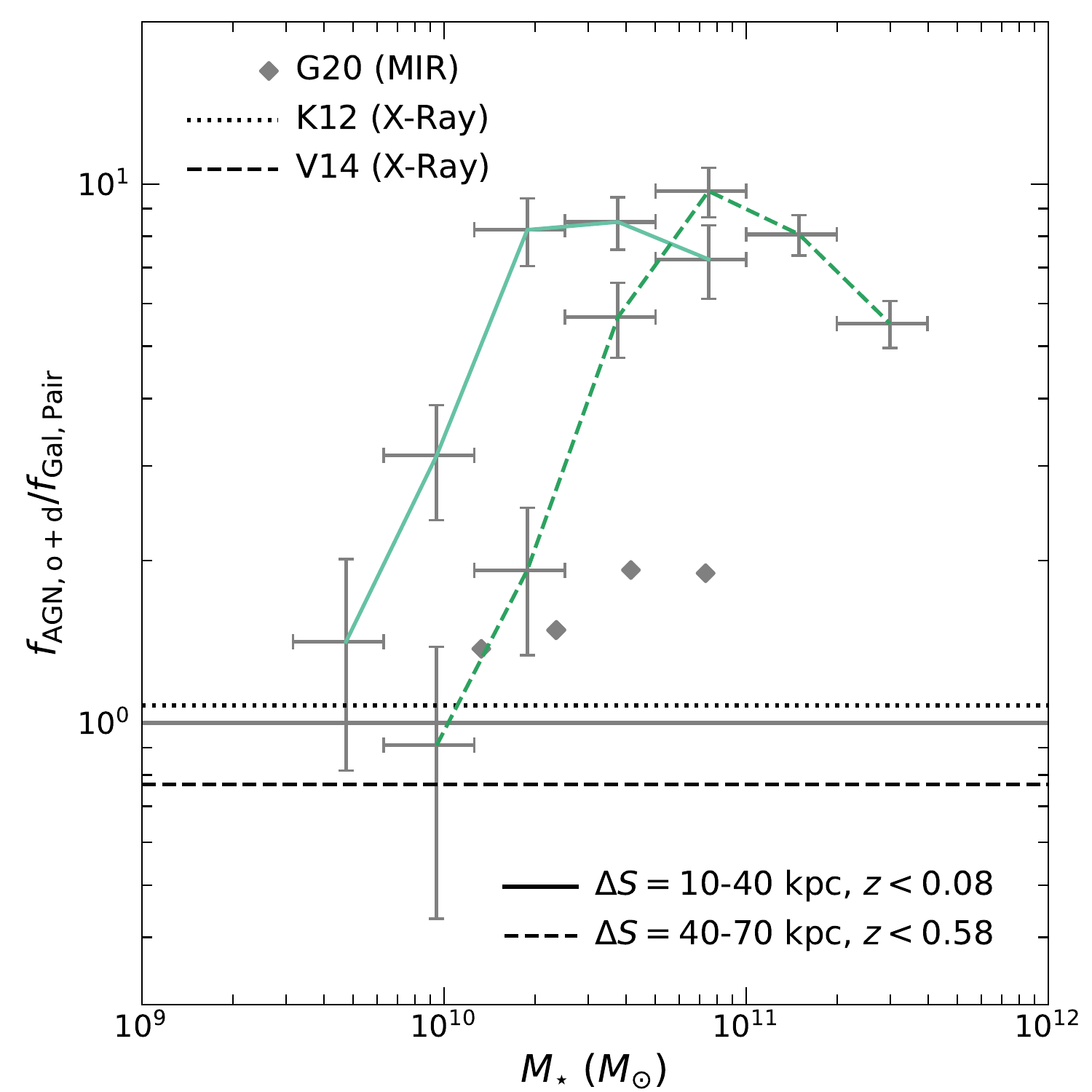}
\caption{\footnotesize{Ratio of \fagnod~to \fgalpair~(i.e. the far left and far right panels of Figure \ref{fig:AGN_GAL_PAIR_FRAC}) as a function of host galaxy stellar mass (\Mstar). The ratios are plotted over the full overlapping \Mstar~range of the samples. The linestyles denoting the redshift and projected physical separation restrictions are the same as in Figure \ref{fig:AGN_GAL_PAIR_FRAC}. The horizontal solid, gray line denotes \fagnod\,$=$\,\fgalpair. The values of \fagnod~and \fgalpair~are comparable at \Mstar\,$\sim$\,$10^{10}$\,\MSun, but their ratio increases to $\sim$10 for \Mstar\,$\sim$\,$10^{11}$\,\MSun. For comparison, gray diamonds denote ratios from a MIR-selected AGN sample (\sdss~sample of \citealt{Gao:2020}; G20), and dotted and dashed lines denote X-ray-selected samples (`Disturbed I' sample from \citealt{Kocevski:2012}; K12 and `Asymmetry' sample from \citealt{Villforth:2014}; V14, respectively). This comparison suggests a stronger preference for MIR-selected AGN to reside in mergers compared to X-ray-selected AGN.}}
\label{fig:AGN_GAL_PAIR_FRAC_EXCESS}
\end{figure}

\begin{figure}[t!]
\includegraphics[width=0.48\textwidth]{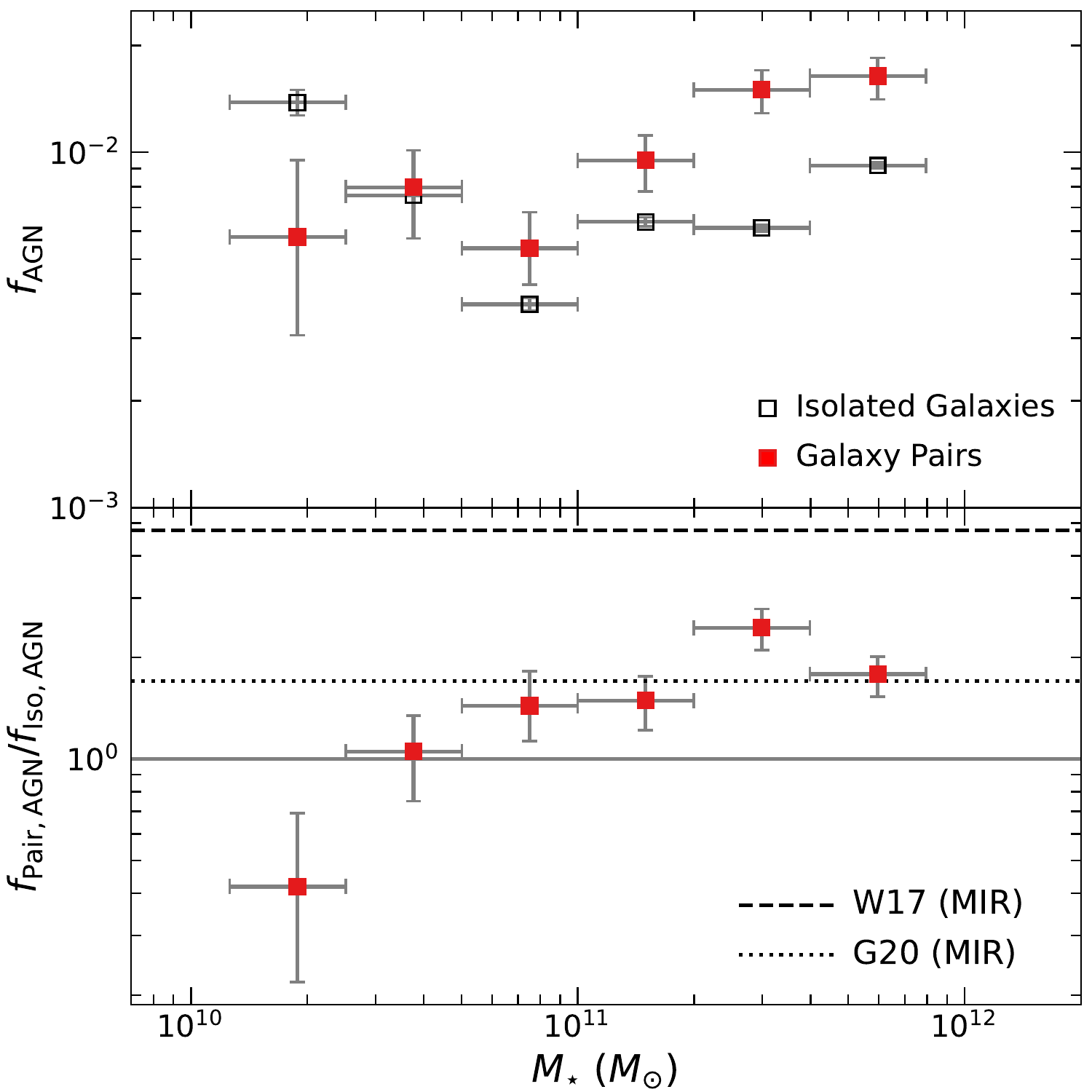}
\caption{\footnotesize{Top: Fraction of isolated galaxies with AGN (\fisoagn, based on \DeltaS\,$>$\,100\,kpc; open squares) versus the fraction of galaxy pairs with AGN (\fmergeragn, from offset and dual AGN systems; filled red squares) as a function of host galaxy stellar mass (\Mstar). For pairs, the value of \Mstar~corresponds to the more massive galaxy. Bottom: Excess of AGN in galaxy pairs (\fmergeragn$/$\fisoagn). The solid gray line denotes \fmergeragn\,$=$\,\fisoagn, with the AGN in pairs exhibiting an excess that generally increases toward higher values of \Mstar. For comparison, dotted and dashed lines denote excesses from MIR-selected AGN samples (\sdss~sample of \citealt{Gao:2020}; G20 and the `Interacting Pairs' sample of \citet{Weston:2017}; W17, respectively). Each MIR-selected sample shows evidence for AGN to preferentially reside in mergers.}}
\label{fig:AGN_MERGER_AGN_GAL_ALL}
\end{figure}

To quantify the AGN merger fraction excess relative to the galaxy merger fraction and compare with previous results, Figure \ref{fig:AGN_GAL_PAIR_FRAC_EXCESS} shows the ratio of \fagnod~to \fgalpair. While \fagnod$/$\fgalpair~is near unity for galaxy stellar masses \Mstar\,$\sim$\,$10^{10}$\,\MSun, it increases toward higher masses, reaching a value of $\sim$10 for \Mstar\,$\sim$\,$10^{11}$\,\MSun. This result suggests that, among massive galaxies, the preference for \wise-selected AGN to reside in galaxy mergers is stronger than that of galaxies without AGN. These excesses generally reflect those from previous MIR AGN studies \citep[e.g.][]{Weston:2017,Gao:2020} and are also generally larger than those using X-ray AGN \citep[e.g.][]{Kocevski:2012,Villforth:2014}. While this trend is qualitatively similar for both the \DeltaS\,$=$\,10\,$-$\,40\,kpc and \DeltaS\,$=$\,40\,$-$\,70\,kpc subsamples, the excesses are systematically larger for the smaller separation pairs (the effect of pair separation is explicitly examined in Section \ref{sec:stage}).

AGN variability on timescales of $<$\,100\,Myr may affect the observed AGN fractions \citep[see][]{Hickox:2014}. Therefore, as an alternative test, we first select galaxy pairs and isolated galaxies (\DeltaS\,$>$\,100\,kpc) and subsequently compare the fraction of isolated galaxies that host an AGN (\fisoagn) to the fraction of galaxy pairs with an AGN (in either offset or dual systems; \fmergeragn). The ratio \fmergeragn$/$\fisoagn~increases with increasing \Mstar, exceeding unity at the largest host masses examined (Figure \ref{fig:AGN_MERGER_AGN_GAL_ALL}). As in Figure \ref{fig:AGN_GAL_PAIR_FRAC_EXCESS}, this excess is similar to that seen in other MIR-selected AGN samples \citep[e.g.][]{Weston:2017,Goulding:2018,Gao:2020}, with quantitative differences possibly due to the variation of merger selection techniques used.

\begin{figure*}[t!]
\includegraphics[width=0.96\textwidth]{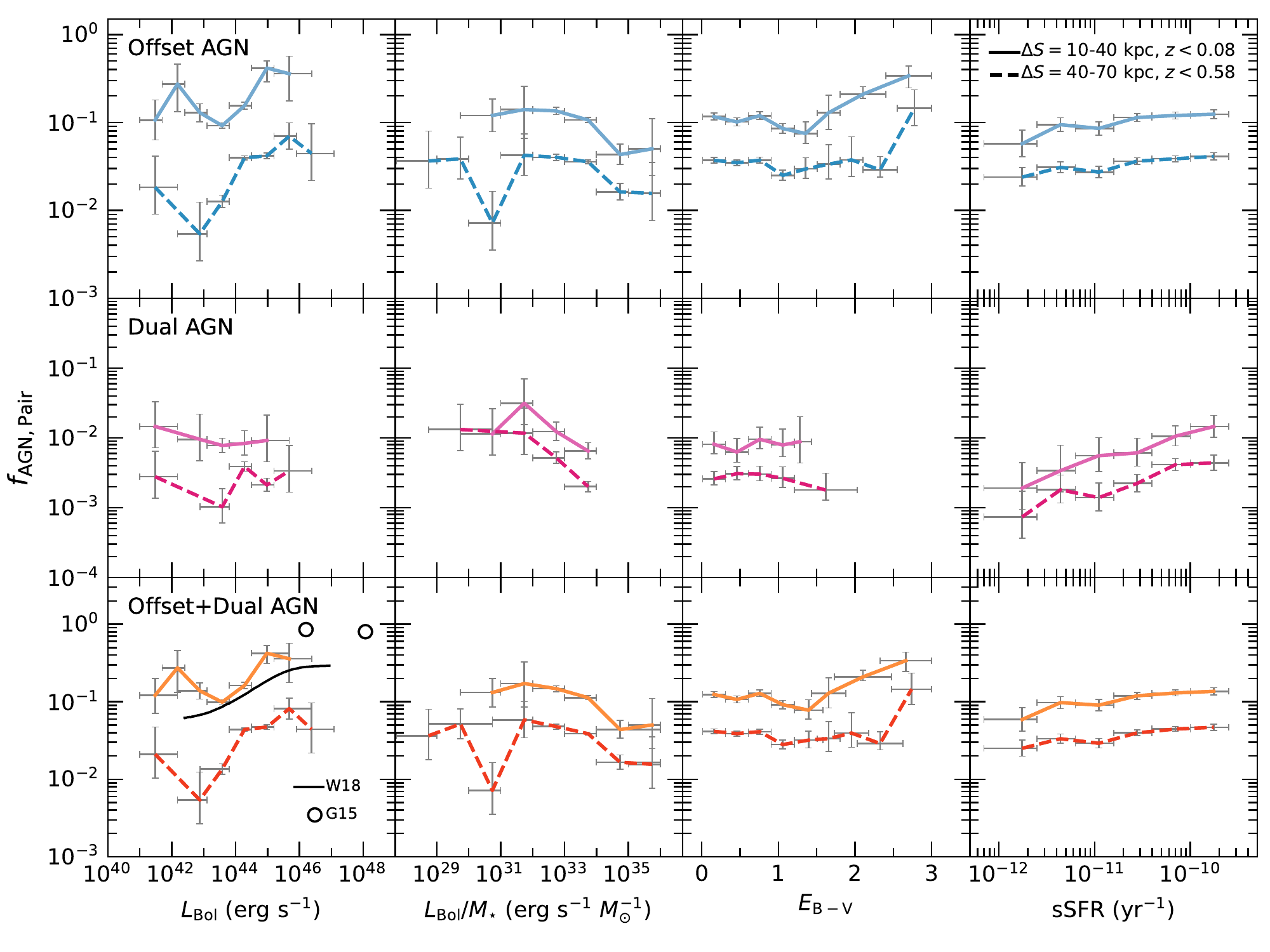}
\caption{\footnotesize{Fraction of unique \wise~AGN in galaxy pairs (\fagnmerger) as a function of (from left to right) AGN bolometric luminosity (\LBol), \LBol~normalized by the host galaxy stellar mass (\LBol\,$/$\,\Mstar), AGN color excess (\EBV, where larger values correspond to higher AGN obscuration), and specific star formation rate (sSFR) for the offset AGN (top), dual AGN (middle), and offset plus dual AGN (bottom). These results are shown for two subsamples in bins of \DeltaS\,$=$\,10\,$-$\,40\,kpc (solid lines) and \DeltaS\,$=$\,40\,$-$\,70\,kpc (dashed lines), limited to $z$\,$\le$\,0.08 and $z$\,$\le$\,0.58 to uniformly resolve down to 10\,kpc and 40\,kpc, respectively, using the 6$''$ resolution of the \wise~source identifications.  For comparison, in the lower left panel, the black solid line denotes a phenomenological model prediction (optimal model from \citealt{Weigel:2018}; W18), and the open black circles denote dust-reddened quasi-stellar objects (\citealt{Glikman:2015}; G15). Positive increases of \fagnmerger~with \LBol~are observed, but essentially vanish when normalized by \Mstar. However, significant positive increases with both \EBV~and sSFR are observed among both offset and dual AGN. This suggests that a significant fraction of merger-driven SMBH growth is obscured for systems with either one or two AGN and may be connected with evolution of the host galaxy stellar populations.}}
\label{fig:LBOL_LEDD_PAIR_FRAC}
\end{figure*}

\section{The Role of Galaxy Mergers for Triggering MIR AGN and SMBH-Galaxy Co-Evolution}
\label{sec:coevol}

In this section we explore how the fractions of AGN in mergers evolve with physical properties that affect the growth of SMBHs and co-evolution with their host galaxies. Specifically, in Section \ref{sec:accretion} we examine the dependence on AGN bolometric luminosity, in Section \ref{sec:obscuration} we examine the dependence on obscuration and host galaxy star formation, and in Section \ref{sec:location} we determine which galaxies of pairs host the more rapidly growing SMBHs.

\subsection{Galaxy Mergers Do Not Enhance Accretion Efficiencies of WISE-Selected AGN}
\label{sec:accretion}

Several works have shown that the AGN merger fraction increases with AGN luminosity among samples selected from optical, X-ray, and MIR observations \citep[e.g.][]{Treister:2012,Barrows:2017} and that the most luminous AGN are perhaps preferentially associated with on-going galaxy interactions and post-coalescence systems \citep[e.g.][]{Urrutia:2008,Glikman:2015}. With the intrinsic AGN bolometric luminosities available through our SED models, in the left panel of Figure \ref{fig:LBOL_LEDD_PAIR_FRAC} we show how the AGN pair fractions of our sample evolve with \LBol. Overall, an increase with \LBol~is observed among the offset and dual AGN samples, consistent with predictions from the phenomenological model of \citet{Weigel:2018}. This trend is similar for both the \DeltaS\,$=$\,10\,$-$\,40\,kpc and \DeltaS\,$=$\,40\,$-$\,70\,kpc subsamples (as with the dependence on \Mstar, the fractions are higher for the smaller separation pairs). The trends show potential evidence for leveling off below \LBol\,$\sim$\,$10^{42}$\,erg\,s$^{-1}$, possibly indicating that low luminosity AGN triggering is mostly independent of mergers. Also shown are the disturbed fractions for luminous dust-reddened quasi-stellar objects from \citet{Glikman:2015} that are consistent with an extrapolation of our \DeltaS\,$=$\,10\,$-$\,40\,kpc subsample out to comparably high luminosities.

However, some numerical and analytical simulations predict that the observed increase of the AGN merger fraction with luminosity is primarily driven by host galaxy stellar mass \citep{Steinborn:2016,Weigel:2018}. These predictions suggest that the Eddington ratios of AGN in galaxy mergers are not significantly different from those of the general AGN population. On the other hand, recent analysis of different simulation results does suggest merger-driven enhancements of Eddington ratios \citep{McAlpine:2020,Byrne-Mamahit:2022}. With our measurements of host galaxy stellar mass that account for the AGN contribution, we test these predictions by examining the mass-normalized values of \LBol~(\LBol\,$/$\,\Mstar) in the second panel of Figure \ref{fig:LBOL_LEDD_PAIR_FRAC}. This normalization effectively removes the positive correlations observed with \LBol: $<$\,1$\sigma$ significance of a non-zero slope (positive or negative, based on a powerlaw fit) is seen among the offset and dual AGN, regardless of pair separation.  Under the assumption that the SMBH mass scales with \Mstar, the AGN merger fractions do not increase with Eddington ratios, and mergers do not enhance the accretion efficiencies of \wise-selected AGN. This result aligns with predictions that suggest the luminosities of AGN in mergers are still dominated by the properties of their host galaxies (e.g. stellar mass and cold gas supplies).  While the potential negative trend with \LBol\,$/$\,\Mstar~is not significant, if real it could be reflecting the bias for AGN in mergers to reside in more massive host galaxies (i.e. Section \ref{sec:agn_gal_mstar_ssfr} and Figure \ref{fig:AGN_GAL_PAIR_FRAC}), corresponding to lower Eddington ratios.

\subsection{Galaxy Mergers Enhance AGN Obscuration and Star Formation Efficiencies}
\label{sec:obscuration}

Simulations predict that a significant fraction of merger-driven SMBH growth is obscured \citep[e.g.][]{Blecha:2017}. However, observational tests of such predictions have been limited by biases toward unobscured AGN. In particular, while connections between mergers and obscuration have been measured using X-ray selected AGN \citep[e.g.][]{Ricci:2017,Ricci:2021}, many heavily obscured sources are still missed (see Section \ref{sec:intro}). Moreover, AGN samples selected using MIR diagnostics are not necessarily obscured, and MIR photometry alone is not sufficient to determine the level of obscuration \citep[e.g.][]{Hickox:2007,Netzer:2015}. Additionally, relying simply on MIR\,$-$\,optical colors to infer the level of AGN obscuration can introduce complications from star formation contributions. The sample developed through this work overcomes these limitations by measuring AGN obscuration from combined AGN and galaxy broadband SED models.

The third panel of Figure \ref{fig:LBOL_LEDD_PAIR_FRAC} shows the AGN pair fractions as a function of AGN color excess (\EBV). With the large sample size and dynamic range of obscuration measurements, we are able to test this connection over the range \EBV\,$\approx$\,0\,$-$\,3. For the first time, we show how the fraction of AGN in offset and dual systems both increase with increasing \EBV~(i.e. toward higher values of AGN obscuration). The trend is strongest in the \DeltaS\,$=$\,10\,$-$\,40\,kpc subsample, with a best-fit powerlaw slope $>$\,0 at a significance of \SlopeSigebvUSAaa$\sigma$ for the offset and dual AGN combined. Still, the positive trend observed at \EBV\,$\gtrsim$\,2 among the \DeltaS\,$=$\,40\,$-$\,70\,kpc subsample suggests that nuclear obscuration may still be enhanced among larger separation pairs. Indeed, observations show that galaxy pairing out to separations of $\gtrsim$40\,kpc may enhance the fraction of obscured AGN \citep[e.g.][]{Satyapal:2014}. We are not able to test for a similar result among the dual AGN due to smaller sample sizes and hence a limited range of \EBV~values.

These extinction values show that the fraction of AGN in mergers significantly increases (up to $\sim$40\%) among sources with column densities $n_{H}$\,$\gtrsim$\,$10^{23}$\,cm$^{-2}$ (using the gas-to-dust ratio of \citealt{Maiolino:2001}), and particularly for those with smaller physical pair separations. By implication, galaxy interactions do contribute significantly to the population of obscured AGN (removing, to a large degree, the dependence on dusty torus orientations).

\begin{figure}[t!]
\includegraphics[width=0.48\textwidth]{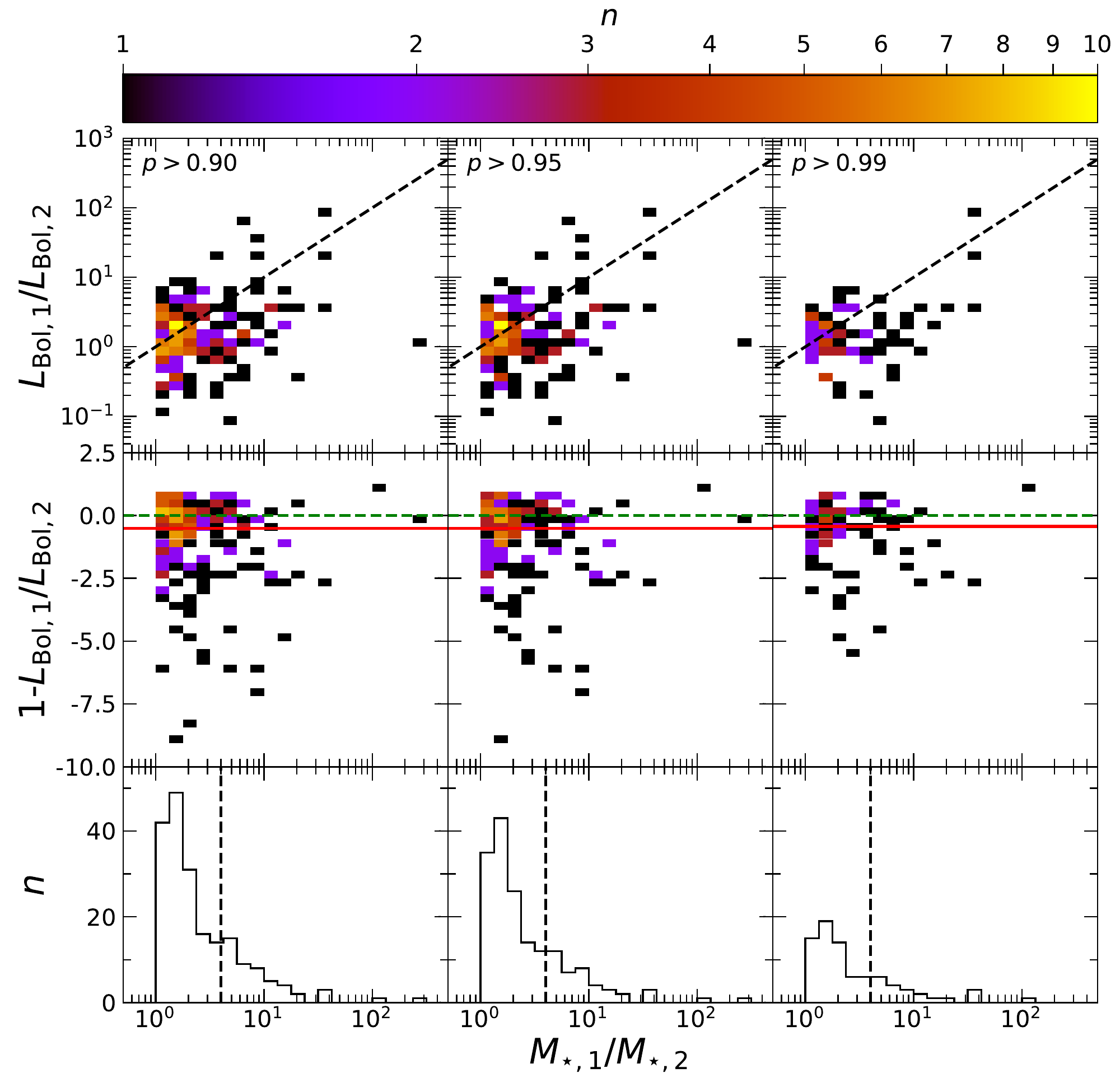}
\caption{\footnotesize{Top: Density plots of the AGN bolometric luminosity ratio (\LBolOne\,$/$\,\LBolTwo) against the host galaxy stellar mass ratio (\MstarOne\,$/$\,\MstarTwo) where the subscripts `1' and `2' denote the more and less massive galaxy, respectively. These plots are shown for dual AGN with \DeltaS\,$\le$\,100\,kpc (based on the redshift of the maximum pair probability) and with total pair probabilities of $p$\,$>$\,90\%~(left), $p$\,$>$\,95\%~(middle), and $p$\,$>$\,99\%~(right). The black dashed line indicates the one-to-one relation: \LBolOne\,$/$\,\LBolTwo\,$=$\,\MstarOne\,$/$\,\MstarTwo. Middle: Offset of \LBolOne\,$/$\,\LBolTwo~from the one-to-one relation shown in the top panels (green dashed line) as a function of \MstarOne\,$/$\,\MstarTwo. The color scale for all panels is shown at the top. For all tested probability thresholds, the median offset (red solid line) is negative due to a tail that extends to low values of \LBolOne\,$/$\,\LBolTwo~per \MstarOne\,$/$\,\MstarTwo. These mergers may be preferentially increasing the Eddington ratios of the AGN in the less massive galaxy. Bottom: Distribution of \MstarOne\,$/$\,\MstarTwo. The black dashed line indicates \MstarOne\,$/$\,\MstarTwo\,$=$\,4 (a commonly adopted division between major and minor mergers), and $\sim$10\%~of the dual AGN are in minor mergers.}}
\label{fig:LBOL_MSTAR}
\end{figure}

AGN and star formation are dependent on a common supply of cold gas and dust, and numerical work suggests this connection is strongest when galaxies are merging due to the enhanced dynamical forces involved \citep[e.g.][]{Volonteri:2015}. Indeed, assuming a direct connection between star-forming galaxies and AGN hosts (when accounting for AGN variability) can naturally explain the observed correlation between the AGN merger fraction and \LBol~\citep[i.e.][]{Hickox:2014}. However, exploring the connection between AGN and star formation requires disentangling their relative contributions. The sample presented here overcomes this limitation with physical properties of the host galaxies that account for the presence of an AGN, allowing us to test this connection for the largest sample of obscured AGN in mergers.

The right panel of Figure \ref{fig:LBOL_LEDD_PAIR_FRAC} shows the fraction of AGN in galaxy pairs as a function of host galaxy specific SFR (sSFR). For \wise~AGN in either offset or dual systems, the pair fractions rise with sSFR. For the combined offset and dual AGN, the best-fit powerlaw slope is $>$\,0 at a significance of \DiffSigssfrUAAaa$\sigma$ in the \DeltaS\,$=$\,10\,$-$\,40\,kpc sample, rising to \DiffSigssfrUAAbb$\sigma$ for the \DeltaS\,$=$\,40\,$-$\,70\,kpc subsample. This systematic increase in AGN pair fraction suggests that the global sSFR may be connected with AGN triggering in mergers (though not necessarily with Eddington ratios; i.e. Section \ref{sec:accretion}). Furthermore, this positive trend may be stronger among the dual AGN (powerlaw exponent of \SlopessfrUAAaa) than among the offset AGN (powerlaw exponent of \SlopessfrUSAaa). However, we do not find any evidence for a significant positive or negative correlation between \LBol~and sSFR among the offset or dual AGN (when limited to a pair probability threshold $p$\,$>$\,90\%), suggesting that a direct connection may be washed out by the different spatial and temporal scales between global star formation and AGN nuclear emission \citep[e.g.][]{Barrows:2017b}.

\begin{figure*}[t!]
\includegraphics[width=0.96\textwidth]{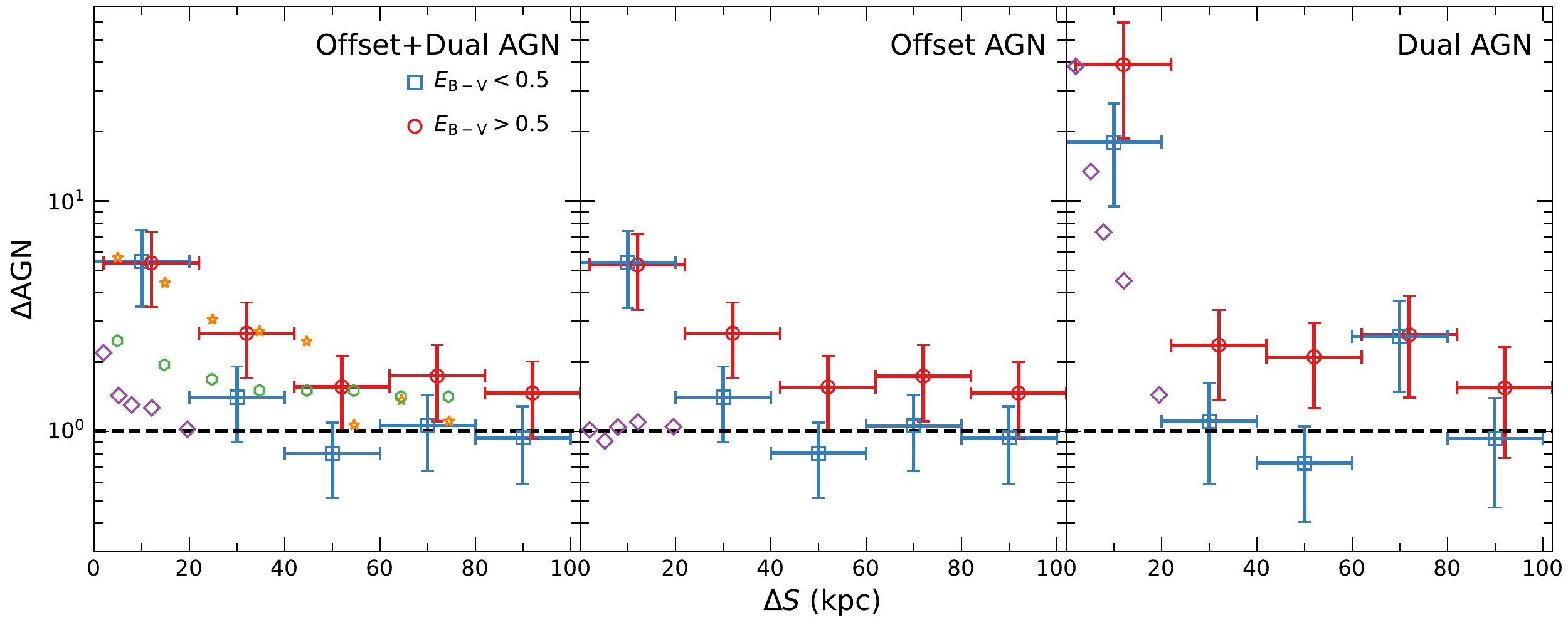}
\caption{\footnotesize{The ratios $N_{\rm{o+dAGN,V}}/N^{'}_{\rm{o+dAGN,V}}$ ($\Delta \rm{o+dAGN}$; left), $N_{\rm{oAGN,V}}/N^{'}_{\rm{oAGN,V}}$ ($\Delta \rm{oAGN}$; middle), and $N_{\rm{dAGN,V}}/N^{'}_{\rm{dAGN,V}}$ ($\Delta \rm{dAGN}$; right) against projected physical separation (\DeltaS). Shown in each panel are the subsets with low AGN obscuration (\EBV\,$<$\,0.5; blue squares) and high AGN obscuration (\EBV\,$>$\,0.5; red circles, horizontally offset for clarity), where (for dual AGN) the values of \EBV~correspond to the AGN in the more massive galaxy. Ratios of unity are indicated by the dashed black line. The horizontal errorbars represent the bin width while the vertical errorbars correspond to the upper and lower Poisson confidence intervals. The excesses systematically increase toward small \DeltaS~in both subsets, and they are higher among the dual AGN. Moreover, the excesses are systematically elevated when limited to the obscured subset. For comparison, the excesses from optical samples (green hexagons: \citealp{Ellison:2013} and purple diamonds: \citealp{Steffen:2022}) and a MIR sample (orange stars: \citealp{Satyapal:2014}) are also shown. The MIR AGN merger excesses are elevated relative to those of lower luminosity optically-selected AGN for \DeltaS\,$<$\,20\,kpc.}}
\label{fig:VOL_DENS_EXCESS}
\end{figure*}

\subsection{Galaxy Mergers May Preferentially Grow SMBHs in the Less Massive Galaxy}
\label{sec:location}

The results from Section \ref{sec:agn_gal_mstar_ssfr} suggest that galaxy mergers can significantly enhance the rate of AGN triggering. In phenomenological and numerical models, this process may result in rapid growth of the SMBH in the less massive galaxy, exceeding that of the SMBH in the primary galaxy \citep[e.g.][]{Yu:2011,Capelo:2016}. If so, unequal mass mergers may be a significant driver of low-mass black hole growth \citep[e.g.][]{Barrows:2018,Barrows:2019}. The sample of dual AGN developed through this work is ideal for testing this scenario since the host galaxy of each AGN is known, along with the host stellar mass and AGN bolometric luminosity, allowing for an observational test of the physics that drives accretion onto SMBHs in galaxy mergers.

Assuming the SMBH mass scales directly with \Mstar, we examine the relative Eddington ratios of both AGN in the dual AGN systems by plotting \LBolOne$/$\LBolTwo~against \MstarOne$/$\MstarTwo, where the subscripts `1' and `2' denote the more and less massive galaxies, respectively (top panels of Figure \ref{fig:LBOL_MSTAR}). Since this analysis assumes each plotted pair is a true pair, we limit it to those with pair probabilities $p$\,$>$\,90\%, and we examine the effect of increasingly stricter criteria ($p$\,$>$\,95\%~and $p$\,$>$\,99\%). To quantify the deviation from the one-to-one relation, the middle panels of Figure \ref{fig:LBOL_MSTAR} show the offset of \LBolOne$/$\LBolTwo~from the one-to-one relation as a function of \MstarOne$/$\MstarTwo. Regardless of probability threshold, the median offset is negative. This offset is due to a significant tail of pairs in which \LBol$/$\Mstar~is larger in the less massive galaxy. 

Since the fraction of dual AGN with negative offsets of \LBolOne$/$\LBolTwo~from the one-to-one relation increases with \MstarOne$/$\MstarTwo, this preferential growth may most likely occur among minor mergers. When adopting a division between major and minor mergers of \MstarOne$/$\MstarTwo\,$=$\,4, $\sim$10\%~of the dual AGN are in minor mergers (bottom panels of Figure \ref{fig:LBOL_MSTAR}). Furthermore, among the offset AGN, the stellar mass ratios between the AGN host and companion galaxy have a median value of $\sim$0.75 (for the same pair probability thresholds used for the dual AGN). These combined results suggest that the mergers may be funneling comparable amounts of gas and dust to each nucleus, thereby preferentially increasing the Eddington ratios of the SMBHs in the less massive galaxies.

\section{Obscured AGN Are Preferentially Found in Late-Stage Dual Systems}
\label{sec:stage}

\begin{figure}[t!]
\includegraphics[width=0.48\textwidth]{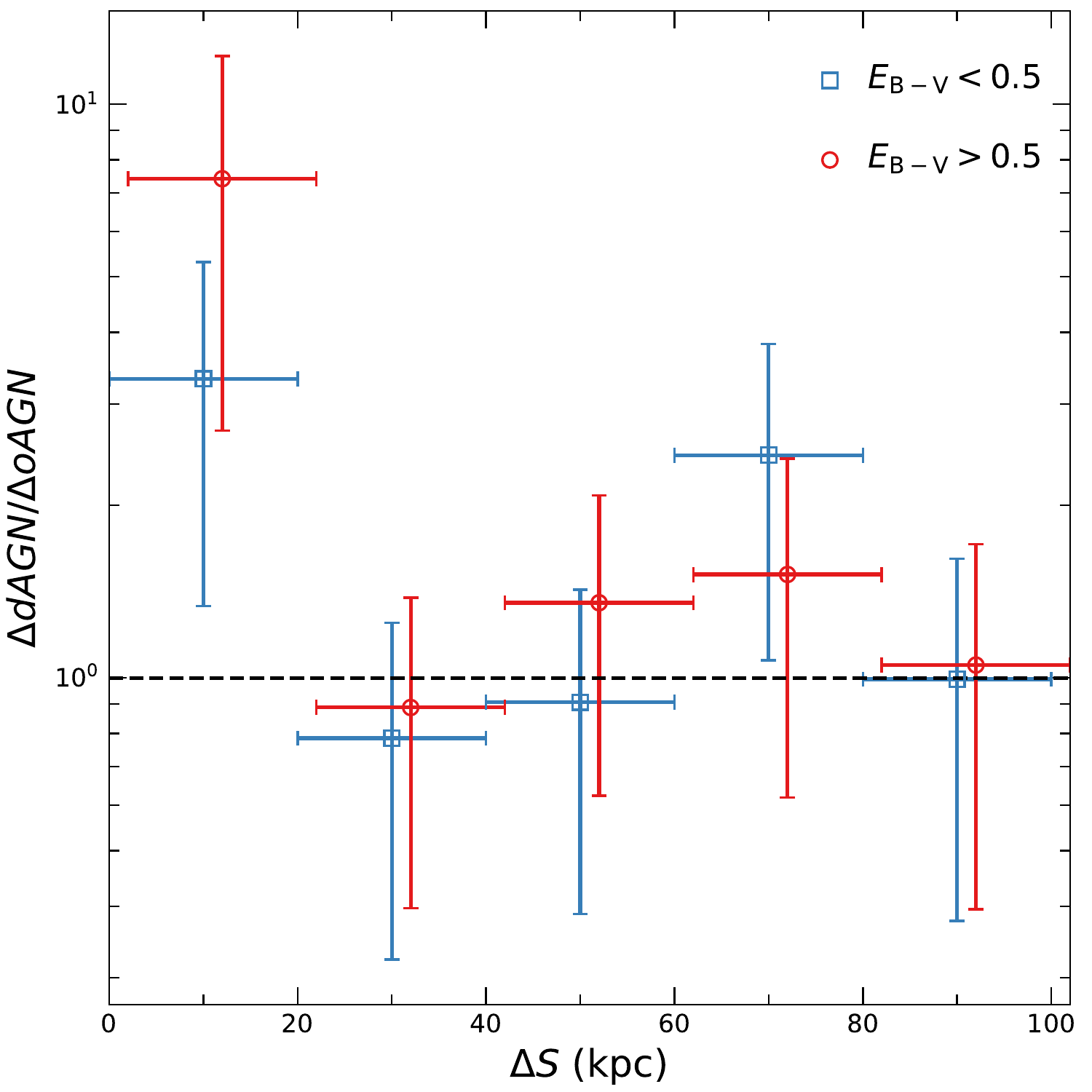}
\caption{\footnotesize{Ratio of the dual and offset AGN excesses ($\Delta \rm{dAGN}$/$\Delta \rm{oAGN}$) from Figure \ref{fig:VOL_DENS_EXCESS} as a function of projected physical separation (\DeltaS) for the unobscured (\EBV\,$<$\,0.5; blue squares) and obscured (\EBV\,$>$\,0.5; red circles, horizontally offset for clarity) subsets. A ratio of unity is indicated by the dashed black line. An increase in the dual AGN excess, relative to the offset AGN excess, is observed toward smaller values of \DeltaS.}}
\label{fig:VOL_DENS_EXCESS_RATIO}
\end{figure}
 
Numerical simulations consistently predict that the late stages of galaxy mergers are when the most significant merger-driven AGN triggering and SMBH accretion occurs \citep[e.g.][]{Hernquist:1989,Mihos:1996,Hopkins2008,Van_Wassenhove:2012,Stickley:2014,Capelo:2015,Blecha:2017}. Hence, observed AGN enhancements in mergers are expected to be strongest at small pair separations. Indeed, previous results have shown that the AGN merger fraction (for X-ray and/or MIR-selected AGN) increases toward smaller pair separations \citep[e.g.][]{Barrows:2017,Stemo:2021}. Moreover, comparison to control samples of inactive galaxies have generally revealed that mergers have an excess of AGN at increasingly later merger stages \citep[e.g.][]{Ellison:2011,Ellison:2013,Satyapal:2014,Fu:2018,Shah:2020,Steffen:2022}. However, these results are limited to systems where offset versus dual AGN can not be distinguished, or otherwise to nearby optically-selected samples for which the dependence on AGN obscuration can not be determined. To bridge this gap, we use our samples of offset and dual AGN, with measures of obscuration, to examine the enhancement of merger-driven AGN in pairs, relative to that expected from stochastic triggering, as a function of \DeltaS.

From the galaxy sample, we first compute the observed number of offset AGN, dual AGN, and galaxy pairs as a function of \DeltaS. To create a volume-limited sample, similar to \citet{Fu:2018} and \citet{Steffen:2022} we normalize the probabilities of each pair by the volume weights ($V_{\rm{max}}$). Doing so yields volume densities for the offset AGN, dual AGN, offset plus dual AGN, and galaxy pairs: \Noagnvmax, \Ndagnvmax, \Nodagnvmax, and \Nggvmax, respectively. Estimates of $V_{\rm{max}}$ are computed at each redshift step of the pair probability distributions by applying the galaxy stellar mass estimate (of the less massive galaxy) at that redshift to the flux-based mass-completeness function described in Section \ref{sec:mass_completeness}.

We then estimate the expected number of AGN among the galaxy pairs from stochastic AGN triggering independent of mergers. These estimates are obtained by computing the fraction of the galaxy sample that host \wise~AGN. Following the procedure from \citet{Fu:2018} and \citet{Steffen:2022}, this fraction is parameterized as a log-normal function of host stellar mass and redshift. This function is then applied to components `a' and `b' of each galaxy pair to determine the probability it hosts a \wise~AGN: \fagna~and \fagnb. The predicted volume density of galaxy pairs with randomly occurring offset AGN is $N^{'}_{\rm{oAGN,V}}=N_{\rm{Gal-Gal},V}\times f_{\rm{AGN,a}}\times (f_{\rm{AGN,b}}-1)+N_{\rm{Gal-Gal},V}\times f_{\rm{AGN,b}}\times (f_{\rm{AGN,a}}-1)$. The predicted volume density with randomly occurring dual AGN is $N^{'}_{\rm{dAGN,V}}=N_{\rm{Gal-Gal},V}\times f_{\rm{AGN,a}}\times f_{\rm{AGN,b}}$, and the predicted volume density with either is $N^{'}_{\rm{o+dAGN,V}}=N^{'}_{\rm{oAGN,V}}+N^{'}_{\rm{dAGN,V}}$. We also derive these volume densities for the unobscured (\EBV\,$<$\,0.5) and obscured (\EBV\,$>$\,0.5) subsets.

The ratios $N_{\rm{o+dAGN,V}}/N^{'}_{\rm{o+dAGN,V}}$, $N_{\rm{oAGN,V}}/N^{'}_{\rm{oAGN,V}}$, and $N_{\rm{dAGN,V}}/N^{'}_{\rm{dAGN,V}}$ as a function of \DeltaS~are shown in Figure \ref{fig:VOL_DENS_EXCESS}. The ratios increase toward smaller separations for both offset and dual AGN, implying that the merger process itself triggers \wise~AGN in both types of systems and that this is increasingly more likely at later merger stages (in agreement with the works cited above). 

Both the unobscured and obscured subsets show qualitatively similar trends. However, the excesses of both the offset and dual AGN are systematically larger for the obscured subset ($\sim$30\%~higher than for unobscured AGN). This result suggests that merger-driven triggering of AGN preferentially occurs in heavily obscured nuclei, regardless of whether or not correlated triggering occurs. As also seen in Figure \ref{fig:VOL_DENS_EXCESS}, the \wise~AGN merger excesses are larger than those of nearby optically-selected AGN. While this has been suggested previously based on comparison of MIR and optical AGN selection \citep[i.e.][]{Satyapal:2014}, our results explicitly show how this is dependent on AGN color excess and that this occurs among both dual and offset AGN. Indeed, in contrast to the null excess observed for the optically-selected offset AGN, the \wise~offset AGN excesses do increase toward small \DeltaS~(with excesses up to a factor of $\sim$5). Moreover, this excess relative to optically-selected AGN holds regardless of \wise~AGN obscuration level. Hence, detection of obscured sources may not completely account for the discrepancies between the optical and MIR-selected samples previously observed \citep[e.g.][]{Ellison:2013,Satyapal:2014}, and the more luminous nature of \wise~AGN (relative to nearby optical AGN) may also contribute to the stronger merger dependence (as suggested by the luminosity dependence in Figure \ref{fig:LBOL_LEDD_PAIR_FRAC}).

Figure \ref{fig:VOL_DENS_EXCESS_RATIO} shows the ratio of the dual AGN excess to the offset AGN excess as a function of \DeltaS~for the unobscured and obscured subsets. The relatively stronger excess among the dual AGN at small \DeltaS~(factors of $\sim$3 and 7 for obscured and unobscured AGN, respectively) can be explained by correlated triggering, i.e. that the likelihood of one AGN being triggered increases the likelihood of the other being triggered during late merger stages. This distinction between offset and dual AGN was also shown in \citet{Fu:2018} and \citet{Steffen:2022} for nearby optically-selected AGN (based on narrow emission line diagnostics), and here we show it exists for relatively more luminous \wise~AGN as well. For direct comparison, we compute the separation-dependent correlated fraction as in \citet{Steffen:2022}. This procedure estimates that 18\%~of pairs with only one AGN become dual AGN due to correlated triggering (similar to, but larger than, the fraction of 11.5\%~for optically-selected AGN from \citealt{Steffen:2022}). Finally, the probability of correlated AGN triggering at small \DeltaS~may be strongest among obscured AGN, though the current uncertainties preclude this result from being significant.

\section{Conclusions}
\label{sec:conc}

We compute photometric redshift PDFs for a wide-area sample of \wise-selected AGN and a comparison sample of galaxies. From these PDFs we generate probabilities for pairs within 100 kpc projected separations, producing the largest sample of \wise-selected AGN in galaxy pairs (\NAGz~offset AGN and \NAAz~dual AGN; Figures \ref{fig:Pairs_AA} and \ref{fig:Pairs_SA}). With uniform physical properties measured for the AGN and host galaxies of each system, we address several outstanding questions regarding the role of galaxy mergers for driving and obscuring SMBH growth. We examine the fraction of \wise~AGN found in offset and dual AGN systems, compare it to that of galaxies, and quantify its dependence on several parameters that influence AGN triggering in galaxy mergers. Our conclusions are as follows:

\begin{enumerate}

\item \wise-selected AGN show a preference for residing in interacting galaxy pairs (offset or dual AGN), with this preference becoming stronger for larger host stellar masses (up to a factor of $\sim$10; Figures \ref{fig:AGN_GAL_PAIR_FRAC_EXCESS} and \ref{fig:AGN_MERGER_AGN_GAL_ALL}). The number of AGN in dual systems is $\sim$10\%~the number in offset systems, so this merger excess is dominated by galaxy pairs with only one AGN.

\item The \wise~AGN merger fraction does not show a statistically significant increase with mass-normalized AGN bolometric luminosity (Figure \ref{fig:LBOL_LEDD_PAIR_FRAC}). Hence, while mergers enhance the frequency of AGN triggering, they do not necessarily enhance the Eddington ratios of \wise~AGN.

\item The \wise~AGN merger fraction does increase significantly with AGN obscuration (significance of \SlopeSigebvUSAaa$\sigma$ for the combined offset and dual AGN; Figure \ref{fig:LBOL_LEDD_PAIR_FRAC}). The strongest increase (up to an AGN fraction of $\sim$40\%) occurs for column densities $n_{H}$\,$\gtrsim$\,$10^{23}$\,cm$^{-2}$. This is the first result to show how the AGN merger fraction explicitly evolves with AGN obscuration for both offset and dual AGN systems, and it suggests that a significant fraction of heavily obscured AGN are in galaxy pairs. 

\item The fraction of AGN in pairs increases with host galaxy specific star formation rate (significance of up to \DiffSigssfrUAAbb$\sigma$ for the combined offset and dual AGN; Figure \ref{fig:LBOL_LEDD_PAIR_FRAC}). This enhancement potentially suggests that AGN triggering is correlated with star formation in mergers and that these systems are potential sites of merger-driven co-evolution between galaxies and SMBHs.

\item For dual AGN systems, the AGN in the less massive galaxies are potentially biased toward larger Eddington ratios compared to those in the more massive galaxies (Figure \ref{fig:LBOL_MSTAR}). Mergers may therefore be effective at growing relatively low-mass SMBHs.

\item The incidence of merger-driven AGN triggering increases toward smaller pair separations (up to an excess of $\sim$5 relative to stochastic mechanisms) and is stronger than that of optically-selected AGN for both offset and dual systems (Figure \ref{fig:VOL_DENS_EXCESS}). This trend is stronger for dual AGN (by a factor of $\sim$3\,$-$\,7; Figure \ref{fig:VOL_DENS_EXCESS_RATIO}), suggesting that the dynamics triggering \wise-selected AGN in mergers also promote correlated AGN triggering in both galaxies.

\item The AGN merger excesses suggest a preference for merger-driven SMBH growth to occur under obscured conditions ($\sim$30\%~higher than for unobscured AGN). Moreover, this dependence on obscuration at late merger stages may be strongest among dual AGN. However, even unobscured \wise~AGN merger excesses are larger than those of lower luminosity optically-selected AGN, suggesting that the luminous nature of \wise~AGN also contributes to their strong merger dependence.

\end{enumerate}

\acknowledgments
{We thank an anonymous reviewer for the detailed and insightful comments that have greatly improved the manuscript quality. This work is supported by the NASA Astrophysics Data Analysis Program 18-ADAP18-0138. RJA was supported by FONDECYT grant number 1191124 and by ANID BASAL project FB210003. The work of DS was carried out at the Jet Propulsion Laboratory, California Institute of Technology, under a contract with NASA.}

\facilities{\wise, \galex, Sloan, PS1, CTIO:2MASS, FLWO:2MASS}

\software{\LRTtitle\footnote{\href{\lrtlink}{\lrtlink}} \citep{Assef:2008,Assef2010}, \astropy\footnote{\href{\astropylink}{\astropylink}} \citep{astropy:2013, astropy:2018}.} \\


\begin{thebibliography}{}
\expandafter\ifx\csname natexlab\endcsname\relax\def\natexlab#1{#1}\fi
\providecommand{\url}[1]{\href{#1}{#1}}
\providecommand{\dodoi}[1]{doi:~\href{http://doi.org/#1}{\nolinkurl{#1}}}
\providecommand{\doeprint}[1]{\href{http://ascl.net/#1}{\nolinkurl{http://ascl.net/#1}}}
\providecommand{\doarXiv}[1]{\href{https://arxiv.org/abs/#1}{\nolinkurl{https://arxiv.org/abs/#1}}}

\bibitem[{{Aird} {et~al.}(2012){Aird}, {Coil}, {Moustakas}, {Blanton},
  {Burles}, {Cool}, {Eisenstein}, {Smith}, {Wong}, \& {Zhu}}]{Aird:2012}
{Aird}, J., {Coil}, A.~L., {Moustakas}, J., {et~al.} 2012, \apj, 746, 90,
  \dodoi{10.1088/0004-637X/746/1/90}

\bibitem[{{Arnouts} {et~al.}(1999){Arnouts}, {Cristiani}, {Moscardini},
  {Matarrese}, {Lucchin}, {Fontana}, \& {Giallongo}}]{Arnouts:1999}
{Arnouts}, S., {Cristiani}, S., {Moscardini}, L., {et~al.} 1999, \mnras, 310,
  540, \dodoi{10.1046/j.1365-8711.1999.02978.x}

\bibitem[{{Ashby} {et~al.}(2009){Ashby}, {Stern}, {Brodwin}, {Griffith},
  {Eisenhardt}, {Koz{\l}owski}, {Kochanek}, {Bock}, {Borys}, {Brand}, {Brown},
  {Cool}, {Cooray}, {Croft}, {Dey}, {Eisenstein}, {Gonzalez}, {Gorjian},
  {Grogin}, {Ivison}, {Jacob}, {Jannuzi}, {Mainzer}, {Moustakas},
  {R{\"o}ttgering}, {Seymour}, {Smith}, {Stanford}, {Stauffer}, {Sullivan},
  {van Breugel}, {Willner}, \& {Wright}}]{Ashby:2009}
{Ashby}, M.~L.~N., {Stern}, D., {Brodwin}, M., {et~al.} 2009, \apj, 701, 428,
  \dodoi{10.1088/0004-637X/701/1/428}

\bibitem[{{Assef} {et~al.}(2018){Assef}, {Stern}, {Noirot}, {Jun}, {Cutri}, \&
  {Eisenhardt}}]{Assef:2018}
{Assef}, R.~J., {Stern}, D., {Noirot}, G., {et~al.} 2018, \apjs, 234, 23,
  \dodoi{10.3847/1538-4365/aaa00a}

\bibitem[{{Assef} {et~al.}(2008){Assef}, {Kochanek}, {Brodwin}, {Brown},
  {Caldwell}, {Cool}, {Eisenhardt}, {Eisenstein}, {Gonzalez}, {Jannuzi},
  {Jones}, {McKenzie}, {Murray}, \& {Stern}}]{Assef:2008}
{Assef}, R.~J., {Kochanek}, C.~S., {Brodwin}, M., {et~al.} 2008, \apj, 676,
  286, \dodoi{10.1086/527533}

\bibitem[{{Assef} {et~al.}(2010){Assef}, {Kochanek}, {Brodwin}, {Cool},
  {Forman}, {Gonzalez}, {Hickox}, {Jones}, {Le Floc'h}, {Moustakas}, {Murray},
  \& {Stern}}]{Assef2010}
---. 2010, AJ, 713, 970, \dodoi{10.1088/0004-637X/713/2/970}

\bibitem[{{Assef} {et~al.}(2013){Assef}, {Stern}, {Kochanek}, {Blain},
  {Brodwin}, {Brown}, {Donoso}, {Eisenhardt}, {Jannuzi}, {Jarrett}, {Stanford},
  {Tsai}, {Wu}, \& {Yan}}]{Assef:2013}
{Assef}, R.~J., {Stern}, D., {Kochanek}, C.~S., {et~al.} 2013, \apj, 772, 26,
  \dodoi{10.1088/0004-637X/772/1/26}

\bibitem[{{Astropy Collaboration} {et~al.}(2013){Astropy Collaboration},
  {Robitaille}, {Tollerud}, {Greenfield}, {Droettboom}, {Bray}, {Aldcroft},
  {Davis}, {Ginsburg}, {Price-Whelan}, {Kerzendorf}, {Conley}, {Crighton},
  {Barbary}, {Muna}, {Ferguson}, {Grollier}, {Parikh}, {Nair}, {Unther},
  {Deil}, {Woillez}, {Conseil}, {Kramer}, {Turner}, {Singer}, {Fox}, {Weaver},
  {Zabalza}, {Edwards}, {Azalee Bostroem}, {Burke}, {Casey}, {Crawford},
  {Dencheva}, {Ely}, {Jenness}, {Labrie}, {Lim}, {Pierfederici}, {Pontzen},
  {Ptak}, {Refsdal}, {Servillat}, \& {Streicher}}]{astropy:2013}
{Astropy Collaboration}, {Robitaille}, T.~P., {Tollerud}, E.~J., {et~al.} 2013,
  \aap, 558, A33, \dodoi{10.1051/0004-6361/201322068}

\bibitem[{{Astropy Collaboration} {et~al.}(2018){Astropy Collaboration},
  {Price-Whelan}, {Sip{\H o}cz}, {G{\"u}nther}, {Lim}, {Crawford}, {Conseil},
  {Shupe}, {Craig}, {Dencheva}, {Ginsburg}, {VanderPlas}, {Bradley},
  {P{\'e}rez-Su{\'a}rez}, {de Val-Borro}, {Aldcroft}, {Cruz}, {Robitaille},
  {Tollerud}, {Ardelean}, {Babej}, {Bach}, {Bachetti}, {Bakanov}, {Bamford},
  {Barentsen}, {Barmby}, {Baumbach}, {Berry}, {Biscani}, {Boquien}, {Bostroem},
  {Bouma}, {Brammer}, {Bray}, {Breytenbach}, {Buddelmeijer}, {Burke},
  {Calderone}, {Cano Rodr{\'{\i}}guez}, {Cara}, {Cardoso}, {Cheedella},
  {Copin}, {Corrales}, {Crichton}, {D'Avella}, {Deil}, {Depagne}, {Dietrich},
  {Donath}, {Droettboom}, {Earl}, {Erben}, {Fabbro}, {Ferreira}, {Finethy},
  {Fox}, {Garrison}, {Gibbons}, {Goldstein}, {Gommers}, {Greco}, {Greenfield},
  {Groener}, {Grollier}, {Hagen}, {Hirst}, {Homeier}, {Horton}, {Hosseinzadeh},
  {Hu}, {Hunkeler}, {Ivezi{\'c}}, {Jain}, {Jenness}, {Kanarek}, {Kendrew},
  {Kern}, {Kerzendorf}, {Khvalko}, {King}, {Kirkby}, {Kulkarni}, {Kumar},
  {Lee}, {Lenz}, {Littlefair}, {Ma}, {Macleod}, {Mastropietro}, {McCully},
  {Montagnac}, {Morris}, {Mueller}, {Mumford}, {Muna}, {Murphy}, {Nelson},
  {Nguyen}, {Ninan}, {N{\"o}the}, {Ogaz}, {Oh}, {Parejko}, {Parley}, {Pascual},
  {Patil}, {Patil}, {Plunkett}, {Prochaska}, {Rastogi}, {Reddy Janga},
  {Sabater}, {Sakurikar}, {Seifert}, {Sherbert}, {Sherwood-Taylor}, {Shih},
  {Sick}, {Silbiger}, {Singanamalla}, {Singer}, {Sladen}, {Sooley},
  {Sornarajah}, {Streicher}, {Teuben}, {Thomas}, {Tremblay}, {Turner},
  {Terr{\'o}n}, {van Kerkwijk}, {de la Vega}, {Watkins}, {Weaver}, {Whitmore},
  {Woillez}, {Zabalza}, \& {Astropy Contributors}}]{astropy:2018}
{Astropy Collaboration}, {Price-Whelan}, A.~M., {Sip{\H o}cz}, B.~M., {et~al.}
  2018, \aj, 156, 123, \dodoi{10.3847/1538-3881/aabc4f}

\bibitem[{Athanassoula(2003)}]{Athanassoula:2003}
Athanassoula, E. 2003, Monthly Notices of the Royal Astronomical Society, 341,
  1179

\bibitem[{{Barrows} {et~al.}(2018){Barrows}, {Comerford}, \&
  {Greene}}]{Barrows:2018}
{Barrows}, R.~S., {Comerford}, J.~M., \& {Greene}, J.~E. 2018, \apj, 869, 154,
  \dodoi{10.3847/1538-4357/aaedb6}

\bibitem[{{Barrows} {et~al.}(2016){Barrows}, {Comerford}, {Greene}, \&
  {Pooley}}]{Barrows:2016}
{Barrows}, R.~S., {Comerford}, J.~M., {Greene}, J.~E., \& {Pooley}, D. 2016,
  \apj, 829, 37, \dodoi{10.3847/0004-637X/829/1/37}

\bibitem[{{Barrows} {et~al.}(2017{\natexlab{a}}){Barrows}, {Comerford},
  {Greene}, \& {Pooley}}]{Barrows:2017}
---. 2017{\natexlab{a}}, \apj, 838, 129, \dodoi{10.3847/1538-4357/aa64d9}

\bibitem[{{Barrows} {et~al.}(2021){Barrows}, {Comerford}, {Stern}, \&
  {Assef}}]{Barrows:2021}
{Barrows}, R.~S., {Comerford}, J.~M., {Stern}, D., \& {Assef}, R.~J. 2021,
  \apj, 922, 179, \dodoi{10.3847/1538-4357/ac1352}

\bibitem[{{Barrows} {et~al.}(2017{\natexlab{b}}){Barrows}, {Comerford},
  {Zakamska}, \& {Cooper}}]{Barrows:2017b}
{Barrows}, R.~S., {Comerford}, J.~M., {Zakamska}, N.~L., \& {Cooper}, M.~C.
  2017{\natexlab{b}}, \apj, 850, 27, \dodoi{10.3847/1538-4357/aa93de}

\bibitem[{{Barrows} {et~al.}(2019){Barrows}, {Mezcua}, \&
  {Comerford}}]{Barrows:2019}
{Barrows}, R.~S., {Mezcua}, M., \& {Comerford}, J.~M. 2019, \apj, 882, 181,
  \dodoi{10.3847/1538-4357/ab338a}

\bibitem[{{Bianchi} {et~al.}(2008){Bianchi}, {Chiaberge}, {Piconcelli},
  {Guainazzi}, \& {Matt}}]{Bianchi2008}
{Bianchi}, S., {Chiaberge}, M., {Piconcelli}, E., {Guainazzi}, M., \& {Matt},
  G. 2008, \mnras, 386, 105, \dodoi{10.1111/j.1365-2966.2008.13078.x}

\bibitem[{{Blecha} {et~al.}(2017){Blecha}, {Snyder}, {Satyapal}, \&
  {Ellison}}]{Blecha:2017}
{Blecha}, L., {Snyder}, G.~F., {Satyapal}, S., \& {Ellison}, S.~L. 2017, ArXiv
  e-prints.
\newblock \doarXiv{1711.02094}

\bibitem[{{Boquien} {et~al.}(2019){Boquien}, {Burgarella}, {Roehlly}, {Buat},
  {Ciesla}, {Corre}, {Inoue}, \& {Salas}}]{Boquien:2019}
{Boquien}, M., {Burgarella}, D., {Roehlly}, Y., {et~al.} 2019, \aap, 622, A103,
  \dodoi{10.1051/0004-6361/201834156}

\bibitem[{Bournaud \& Combes(2002)}]{Bournaud:2002}
Bournaud, F., \& Combes, F. 2002, Astronomy \& Astrophysics, 392, 83

\bibitem[{{Byrne-Mamahit} {et~al.}(2022){Byrne-Mamahit}, {Hani}, {Ellison},
  {Quai}, \& {Patton}}]{Byrne-Mamahit:2022}
{Byrne-Mamahit}, S., {Hani}, M., {Ellison}, S., {Quai}, S., \& {Patton}, D.
  2022, arXiv e-prints, arXiv:2212.07342.
\newblock \doarXiv{2212.07342}

\bibitem[{{Capelo} {et~al.}(2016){Capelo}, {Dotti}, {Volonteri}, {Mayer},
  {Bellovary}, \& {Shen}}]{Capelo:2016}
{Capelo}, P.~R., {Dotti}, M., {Volonteri}, M., {et~al.} 2016, ArXiv e-prints.
\newblock \doarXiv{1611.09244}

\bibitem[{{Capelo} {et~al.}(2017){Capelo}, {Dotti}, {Volonteri}, {Mayer},
  {Bellovary}, \& {Shen}}]{Capelo:2017}
---. 2017, \mnras, 469, 4437, \dodoi{10.1093/mnras/stx1067}

\bibitem[{{Capelo} {et~al.}(2015){Capelo}, {Volonteri}, {Dotti}, {Bellovary},
  {Mayer}, \& {Governato}}]{Capelo:2015}
{Capelo}, P.~R., {Volonteri}, M., {Dotti}, M., {et~al.} 2015, \mnras, 447,
  2123, \dodoi{10.1093/mnras/stu2500}

\bibitem[{Combes \& Gerin(1985)}]{Combes:1985}
Combes, F., \& Gerin, M. 1985, Astronomy and Astrophysics, 150, 327

\bibitem[{{Comerford} {et~al.}(2015){Comerford}, {Pooley}, {Barrows}, {Greene},
  {Zakamska}, {Madejski}, \& {Cooper}}]{Comerford:2015}
{Comerford}, J.~M., {Pooley}, D., {Barrows}, R.~S., {et~al.} 2015, \apj, 806,
  219, \dodoi{10.1088/0004-637X/806/2/219}

\bibitem[{{Cunha} {et~al.}(2009){Cunha}, {Lima}, {Oyaizu}, {Frieman}, \&
  {Lin}}]{Cunha:2009}
{Cunha}, C.~E., {Lima}, M., {Oyaizu}, H., {Frieman}, J., \& {Lin}, H. 2009,
  \mnras, 396, 2379, \dodoi{10.1111/j.1365-2966.2009.14908.x}

\bibitem[{{Darvish} {et~al.}(2015){Darvish}, {Mobasher}, {Sobral}, {Scoville},
  \& {Aragon-Calvo}}]{Darvish:2015}
{Darvish}, B., {Mobasher}, B., {Sobral}, D., {Scoville}, N., \& {Aragon-Calvo},
  M. 2015, \apj, 805, 121, \dodoi{10.1088/0004-637X/805/2/121}

\bibitem[{{De Rosa} {et~al.}(2022){De Rosa}, {Vignali}, {Severgnini},
  {Bianchi}, {Bogdanovi{\'c}}, {Charisi}, {Guainazzi}, {Haiman}, {Komossa},
  {Paragi}, {Perez-Torres}, {Piconcelli}, {Ducci}, {Parvatikar}, \&
  {Serafinelli}}]{DeRosa:2022}
{De Rosa}, A., {Vignali}, C., {Severgnini}, P., {et~al.} 2022, \mnras,
  \dodoi{10.1093/mnras/stac3664}

\bibitem[{{Di Matteo} {et~al.}(2005){Di Matteo}, {Springel}, \&
  {Hernquist}}]{DiMatteo:2005}
{Di Matteo}, T., {Springel}, V., \& {Hernquist}, L. 2005, \nat, 433, 604,
  \dodoi{10.1038/nature03335}

\bibitem[{{Duncan} {et~al.}(2019){Duncan}, {Conselice}, {Mundy}, {Bell},
  {Donley}, {Galametz}, {Guo}, {Grogin}, {Hathi}, {Kartaltepe}, {Kocevski},
  {Koekemoer}, {P{\'e}rez-Gonz{\'a}lez}, {Mantha}, {Snyder}, \&
  {Stefanon}}]{Duncan:2019}
{Duncan}, K., {Conselice}, C.~J., {Mundy}, C., {et~al.} 2019, \apj, 876, 110,
  \dodoi{10.3847/1538-4357/ab148a}

\bibitem[{{Duncan} {et~al.}(2018){Duncan}, {Brown}, {Williams}, {Best}, {Buat},
  {Burgarella}, {Jarvis}, {Ma{\l}ek}, {Oliver}, {R{\"o}ttgering}, \&
  {Smith}}]{Duncan:2018}
{Duncan}, K.~J., {Brown}, M. J.~I., {Williams}, W.~L., {et~al.} 2018, \mnras,
  473, 2655, \dodoi{10.1093/mnras/stx2536}

\bibitem[{{Ellison} {et~al.}(2013){Ellison}, {Mendel}, {Scudder}, {Patton}, \&
  {Palmer}}]{Ellison:2013}
{Ellison}, S.~L., {Mendel}, J.~T., {Scudder}, J.~M., {Patton}, D.~R., \&
  {Palmer}, M.~J.~D. 2013, \mnras, 430, 3128, \dodoi{10.1093/mnras/sts546}

\bibitem[{{Ellison} {et~al.}(2011){Ellison}, {Patton}, {Mendel}, \&
  {Scudder}}]{Ellison:2011}
{Ellison}, S.~L., {Patton}, D.~R., {Mendel}, J.~T., \& {Scudder}, J.~M. 2011,
  \mnras, 418, 2043, \dodoi{10.1111/j.1365-2966.2011.19624.x}

\bibitem[{{Farrah} {et~al.}(2017){Farrah}, {Petty}, {Connolly}, {Blain},
  {Efstathiou}, {Lacy}, {Stern}, {Lake}, {Jarrett}, {Bridge}, {Eisenhardt},
  {Benford}, {Jones}, {Tsai}, {Assef}, {Wu}, \& {Moustakas}}]{Farrah:2017}
{Farrah}, D., {Petty}, S., {Connolly}, B., {et~al.} 2017, \apj, 844, 106,
  \dodoi{10.3847/1538-4357/aa78f2}

\bibitem[{{Fern{\'a}ndez-Soto} {et~al.}(2002){Fern{\'a}ndez-Soto}, {Lanzetta},
  {Chen}, {Levine}, \& {Yahata}}]{Fernandez_Soto:2002}
{Fern{\'a}ndez-Soto}, A., {Lanzetta}, K.~M., {Chen}, H.~W., {Levine}, B., \&
  {Yahata}, N. 2002, \mnras, 330, 889, \dodoi{10.1046/j.1365-8711.2002.05131.x}

\bibitem[{{Foreman-Mackey} {et~al.}(2013){Foreman-Mackey}, {Hogg}, {Lang}, \&
  {Goodman}}]{Foreman-Mackey:2013}
{Foreman-Mackey}, D., {Hogg}, D.~W., {Lang}, D., \& {Goodman}, J. 2013, \pasp,
  125, 306, \dodoi{10.1086/670067}

\bibitem[{{Fu} {et~al.}(2018){Fu}, {Steffen}, {Gross}, {Dai}, {Isbell}, {Lin},
  {Wake}, {Xue}, {Bizyaev}, \& {Pan}}]{Fu:2018}
{Fu}, H., {Steffen}, J.~L., {Gross}, A.~C., {et~al.} 2018, \apj, 856, 93,
  \dodoi{10.3847/1538-4357/aab364}

\bibitem[{{Gao} {et~al.}(2020){Gao}, {Wang}, {Pearson}, {Gordon}, {Holwerda},
  {Hopkins}, {Brown}, {Bland-Hawthorn}, \& {Owers}}]{Gao:2020}
{Gao}, F., {Wang}, L., {Pearson}, W.~J., {et~al.} 2020, \aap, 637, A94,
  \dodoi{10.1051/0004-6361/201937178}

\bibitem[{{Georgakakis} {et~al.}(2009){Georgakakis}, {Coil}, {Laird},
  {Griffith}, {Nandra}, {Lotz}, {Pierce}, {Cooper}, {Newman}, \&
  {Koekemoer}}]{Georgakakis:2009}
{Georgakakis}, A., {Coil}, A.~L., {Laird}, E.~S., {et~al.} 2009, \mnras, 397,
  623, \dodoi{10.1111/j.1365-2966.2009.14951.x}

\bibitem[{{Gilli} {et~al.}(2007){Gilli}, {Comastri}, \&
  {Hasinger}}]{Gilli:2007}
{Gilli}, R., {Comastri}, A., \& {Hasinger}, G. 2007, \aap, 463, 79,
  \dodoi{10.1051/0004-6361:20066334}

\bibitem[{{Glikman} {et~al.}(2015){Glikman}, {Simmons}, {Mailly}, {Schawinski},
  {Urry}, \& {Lacy}}]{Glikman:2015}
{Glikman}, E., {Simmons}, B., {Mailly}, M., {et~al.} 2015, \apj, 806, 218,
  \dodoi{10.1088/0004-637X/806/2/218}

\bibitem[{{Goulding} {et~al.}(2018){Goulding}, {Greene}, {Bezanson}, {Greco},
  {Johnson}, {Leauthaud}, {Matsuoka}, {Medezinski}, \&
  {Price-Whelan}}]{Goulding:2018}
{Goulding}, A.~D., {Greene}, J.~E., {Bezanson}, R., {et~al.} 2018, \pasj, 70,
  S37, \dodoi{10.1093/pasj/psx135}

\bibitem[{{Guainazzi} {et~al.}(2005){Guainazzi}, {Piconcelli},
  {Jim{\'e}nez-Bail{\'o}n}, \& {Matt}}]{Guainazzi2005}
{Guainazzi}, M., {Piconcelli}, E., {Jim{\'e}nez-Bail{\'o}n}, E., \& {Matt}, G.
  2005, \aap, 429, L9, \dodoi{10.1051/0004-6361:200400104}

\bibitem[{Heller \& Shlosman(1994)}]{Heller:1994}
Heller, C.~H., \& Shlosman, I. 1994, The Astrophysical Journal, 424, 84

\bibitem[{{Hernquist}(1989)}]{Hernquist:1989}
{Hernquist}, L. 1989, \nat, 340, 687, \dodoi{10.1038/340687a0}

\bibitem[{{Hickox} {et~al.}(2014){Hickox}, {Mullaney}, {Alexander}, {Chen},
  {Civano}, {Goulding}, \& {Hainline}}]{Hickox:2014}
{Hickox}, R.~C., {Mullaney}, J.~R., {Alexander}, D.~M., {et~al.} 2014, \apj,
  782, 9, \dodoi{10.1088/0004-637X/782/1/9}

\bibitem[{{Hickox} {et~al.}(2007){Hickox}, {Jones}, {Forman}, {Murray},
  {Brodwin}, {Brown}, {Eisenhardt}, {Stern}, {Kochanek}, {Eisenstein}, {Cool},
  {Jannuzi}, {Dey}, {Brand}, {Gorjian}, \& {Caldwell}}]{Hickox:2007}
{Hickox}, R.~C., {Jones}, C., {Forman}, W.~R., {et~al.} 2007, \apj, 671, 1365,
  \dodoi{10.1086/523082}

\bibitem[{{Hinshaw} {et~al.}(2013){Hinshaw}, {Larson}, {Komatsu}, {Spergel},
  {Bennett}, {Dunkley}, {Nolta}, {Halpern}, {Hill}, {Odegard}, {Page}, {Smith},
  {Weiland}, {Gold}, {Jarosik}, {Kogut}, {Limon}, {Meyer}, {Tucker}, {Wollack},
  \& {Wright}}]{Hinshaw:2013}
{Hinshaw}, G., {Larson}, D., {Komatsu}, E., {et~al.} 2013, The Astrophysical
  Journal Supplement Series, 208, 19, \dodoi{10.1088/0067-0049/208/2/19}

\bibitem[{{Hoaglin} {et~al.}(1983){Hoaglin}, {Mosteller}, \&
  {Tukey}}]{Hoaglin:1983}
{Hoaglin}, D.~C., {Mosteller}, F., \& {Tukey}, J.~W. 1983, {Understanding
  robust and exploratory data anlysis}

\bibitem[{{Hopkins} {et~al.}(2008){Hopkins}, {Hernquist}, {Cox}, \& {Kere{\v
  s}}}]{Hopkins2008}
{Hopkins}, P.~F., {Hernquist}, L., {Cox}, T.~J., \& {Kere{\v s}}, D. 2008,
  \apjs, 175, 356, \dodoi{10.1086/524362}

\bibitem[{{Hou} {et~al.}(2020){Hou}, {Li}, \& {Liu}}]{Hou:2020}
{Hou}, M., {Li}, Z., \& {Liu}, X. 2020, \apj, 900, 79,
  \dodoi{10.3847/1538-4357/aba4a7}

\bibitem[{{Hou} {et~al.}(2022){Hou}, {Li}, {Liu}, {Li}, {Li}, {Wang}, {Wang},
  \& {Ho}}]{Hou:2022}
{Hou}, M., {Li}, Z., {Liu}, X., {et~al.} 2022, arXiv e-prints,
  arXiv:2212.06399.
\newblock \doarXiv{2212.06399}

\bibitem[{{Hudson} {et~al.}(2006){Hudson}, {Reiprich}, {Clarke}, \&
  {Sarazin}}]{Hudson2006}
{Hudson}, D.~S., {Reiprich}, T.~H., {Clarke}, T.~E., \& {Sarazin}, C.~L. 2006,
  \aap, 453, 433, \dodoi{10.1051/0004-6361:20064955}

\bibitem[{{Ilbert} {et~al.}(2006){Ilbert}, {Arnouts}, {McCracken},
  {Bolzonella}, {Bertin}, {Le F{\`e}vre}, {Mellier}, {Zamorani}, {Pell{\`o}},
  {Iovino}, {Tresse}, {Le Brun}, {Bottini}, {Garilli}, {Maccagni}, {Picat},
  {Scaramella}, {Scodeggio}, {Vettolani}, {Zanichelli}, {Adami}, {Bardelli},
  {Cappi}, {Charlot}, {Ciliegi}, {Contini}, {Cucciati}, {Foucaud}, {Franzetti},
  {Gavignaud}, {Guzzo}, {Marano}, {Marinoni}, {Mazure}, {Meneux}, {Merighi},
  {Paltani}, {Pollo}, {Pozzetti}, {Radovich}, {Zucca}, {Bondi}, {Bongiorno},
  {Busarello}, {de La Torre}, {Gregorini}, {Lamareille}, {Mathez}, {Merluzzi},
  {Ripepi}, {Rizzo}, \& {Vergani}}]{Ilbert:2006}
{Ilbert}, O., {Arnouts}, S., {McCracken}, H.~J., {et~al.} 2006, \aap, 457, 841,
  \dodoi{10.1051/0004-6361:20065138}

\bibitem[{{Jarrett} {et~al.}(2011){Jarrett}, {Cohen}, {Masci}, {Wright},
  {Stern}, {Benford}, {Blain}, {Carey}, {Cutri}, {Eisenhardt}, {Lonsdale},
  {Mainzer}, {Marsh}, {Padgett}, {Petty}, {Ressler}, {Skrutskie}, {Stanford},
  {Surace}, {Tsai}, {Wheelock}, \& {Yan}}]{Jarrett:2011}
{Jarrett}, T.~H., {Cohen}, M., {Masci}, F., {et~al.} 2011, \apj, 735, 112,
  \dodoi{10.1088/0004-637X/735/2/112}

\bibitem[{{Jin} {et~al.}(2021){Jin}, {Dai}, {Pan}, {Lin}, {Li}, {Hsieh},
  {Shen}, {Yuan}, {Feng}, {Cheng}, {Xu}, {Huang}, \& {Zhang}}]{Jin:2021}
{Jin}, G., {Dai}, Y.~S., {Pan}, H.-A., {et~al.} 2021, \apj, 923, 6,
  \dodoi{10.3847/1538-4357/ac2901}

\bibitem[{{Kartaltepe} {et~al.}(2010){Kartaltepe}, {Sanders}, {Le Floc'h},
  {Frayer}, {Aussel}, {Arnouts}, {Ilbert}, {Salvato}, {Scoville}, {Surace},
  {Yan}, {Brusa}, {Capak}, {Caputi}, {Carollo}, {Civano}, {Elvis}, {Faure},
  {Hasinger}, {Koekemoer}, {Lee}, {Lilly}, {Liu}, {McCracken}, {Schinnerer},
  {Smol{\v{c}}i{\'c}}, {Taniguchi}, {Thompson}, \& {Trump}}]{Kartaltep:2010}
{Kartaltepe}, J.~S., {Sanders}, D.~B., {Le Floc'h}, E., {et~al.} 2010, \apj,
  709, 572, \dodoi{10.1088/0004-637X/709/2/572}

\bibitem[{{Kocevski} {et~al.}(2012){Kocevski}, {Faber}, {Mozena}, {Koekemoer},
  {Nandra}, {Rangel}, {Laird}, {Brusa}, {Wuyts}, {Trump}, {Koo}, {Somerville},
  {Bell}, {Lotz}, {Alexander}, {Bournaud}, {Conselice}, {Dahlen}, {Dekel},
  {Donley}, {Dunlop}, {Finoguenov}, {Georgakakis}, {Giavalisco}, {Guo},
  {Grogin}, {Hathi}, {Juneau}, {Kartaltepe}, {Lucas}, {McGrath}, {McIntosh},
  {Mobasher}, {Robaina}, {Rosario}, {Straughn}, {van der Wel}, \&
  {Villforth}}]{Kocevski:2012}
{Kocevski}, D.~D., {Faber}, S.~M., {Mozena}, M., {et~al.} 2012, \apj, 744, 148,
  \dodoi{10.1088/0004-637X/744/2/148}

\bibitem[{{Komossa} {et~al.}(2003){Komossa}, {Burwitz}, {Hasinger}, {Predehl},
  {Kaastra}, \& {Ikebe}}]{Komossa2003}
{Komossa}, S., {Burwitz}, V., {Hasinger}, G., {et~al.} 2003, \apjl, 582, L15,
  \dodoi{10.1086/346145}

\bibitem[{{Koss} {et~al.}(2012){Koss}, {Mushotzky}, {Treister}, {Veilleux},
  {Vasudevan}, \& {Trippe}}]{Koss:2012}
{Koss}, M., {Mushotzky}, R., {Treister}, E., {et~al.} 2012, \apjl, 746, L22,
  \dodoi{10.1088/2041-8205/746/2/L22}

\bibitem[{{Koss} {et~al.}(2011){Koss}, {Mushotzky}, {Treister}, {Veilleux},
  {Vasudevan}, {Miller}, {Sanders}, {Schawinski}, \& {Trippe}}]{Koss:2011}
---. 2011, \apjl, 735, L42+, \dodoi{10.1088/2041-8205/735/2/L42}

\bibitem[{{Koss} {et~al.}(2022){Koss}, {Trakhtenbrot}, {Ricci}, {Bauer},
  {Treister}, {Mushotzky}, {Urry}, {Ananna}, {Balokovi{\'c}}, {den Brok},
  {Cenko}, {Harrison}, {Ichikawa}, {Lamperti}, {Lein}, {Mej{\'\i}a-Restrepo},
  {Oh}, {Pacucci}, {Pfeifle}, {Powell}, {Privon}, {Ricci}, {Salvato},
  {Schawinski}, {Shimizu}, {Smith}, \& {Stern}}]{Koss:2022}
{Koss}, M.~J., {Trakhtenbrot}, B., {Ricci}, C., {et~al.} 2022, \apjs, 261, 1,
  \dodoi{10.3847/1538-4365/ac6c8f}

\bibitem[{{Liu} {et~al.}(2011){Liu}, {Shen}, {Strauss}, \& {Hao}}]{Liu:2011}
{Liu}, X., {Shen}, Y., {Strauss}, M.~A., \& {Hao}, L. 2011, \apj, 737, 101,
  \dodoi{10.1088/0004-637X/737/2/101}

\bibitem[{{L{\'o}pez-Sanjuan} {et~al.}(2010){L{\'o}pez-Sanjuan}, {Balcells},
  {P{\'e}rez-Gonz{\'a}lez}, {Barro}, {Gallego}, \&
  {Zamorano}}]{Lopez_Sanjuan:2010}
{L{\'o}pez-Sanjuan}, C., {Balcells}, M., {P{\'e}rez-Gonz{\'a}lez}, P.~G.,
  {et~al.} 2010, \aap, 518, A20, \dodoi{10.1051/0004-6361/201014236}

\bibitem[{{L{\'o}pez-Sanjuan} {et~al.}(2015){L{\'o}pez-Sanjuan}, {Cenarro},
  {Varela}, {Viironen}, {Molino}, {Ben{\'\i}tez}, {Arnalte-Mur}, {Ascaso},
  {D{\'\i}az-Garc{\'\i}a}, {Fern{\'a}ndez-Soto}, {Jim{\'e}nez-Teja},
  {M{\'a}rquez}, {Masegosa}, {Moles}, {Povi{\'c}}, {Aguerri}, {Alfaro},
  {Aparicio-Villegas}, {Broadhurst}, {Cabrera-Ca{\~n}o}, {Castander}, {Cepa},
  {Cervi{\~n}o}, {Crist{\'o}bal-Hornillos}, {Del Olmo}, {Gonz{\'a}lez Delgado},
  {Husillos}, {Infante}, {Mart{\'\i}nez}, {Perea}, {Prada}, \&
  {Quintana}}]{Lopez_Sanjuan:2015}
{L{\'o}pez-Sanjuan}, C., {Cenarro}, A.~J., {Varela}, J., {et~al.} 2015, \aap,
  576, A53, \dodoi{10.1051/0004-6361/201424913}

\bibitem[{Lynden-Bell(1979)}]{Lynden-Bell:1979}
Lynden-Bell, D. 1979, Monthly Notices of the Royal Astronomical Society, 187,
  101

\bibitem[{{Madsen} {et~al.}(2019){Madsen}, {Hickox}, {Bachetti}, {Stern},
  {Gellert}, {Garc{\'\i}a}, {Kara}, {Brandt}, {Krawczynski}, {Lohfink},
  {Brenneman}, {Christensen}, {Middleton}, {Hornstrup}, {Matt}, {Jaodand},
  {Lansbury}, {Ricci}, {Fuerst}, {Ballantyne}, {Walton}, {Fabian}, {Della
  Monica Ferreira}, {Pottschmidt}, {Miller}, {Windt}, {Balokovi{\'c}},
  {Kamraj}, {Wilms}, {Heida}, {Alexander}, {Boorman}, {Wik}, {Vogel},
  {Earnshaw}, {Descalle}, {Civano}, {Fornasini}, {Grindlay}, {Zhang},
  {Hornschemeier}, \& {Craig}}]{Madsen:2019}
{Madsen}, K., {Hickox}, R., {Bachetti}, M., {et~al.} 2019, in Bulletin of the
  American Astronomical Society, Vol.~51, 166

\bibitem[{{Maiolino} {et~al.}(2001){Maiolino}, {Marconi}, {Salvati},
  {Risaliti}, {Severgnini}, {Oliva}, {La Franca}, \& {Vanzi}}]{Maiolino:2001}
{Maiolino}, R., {Marconi}, A., {Salvati}, M., {et~al.} 2001, \aap, 365, 28,
  \dodoi{10.1051/0004-6361:20000177}

\bibitem[{{Mateos} {et~al.}(2012){Mateos}, {Alonso-Herrero}, {Carrera},
  {Blain}, {Watson}, {Barcons}, {Braito}, {Severgnini}, {Donley}, \&
  {Stern}}]{Mateos:2012}
{Mateos}, S., {Alonso-Herrero}, A., {Carrera}, F.~J., {et~al.} 2012, \mnras,
  426, 3271, \dodoi{10.1111/j.1365-2966.2012.21843.x}

\bibitem[{{Mazzarella} {et~al.}(2012){Mazzarella}, {Iwasawa}, {Vavilkin},
  {Armus}, {Kim}, {Bothun}, {Evans}, {Spoon}, {Haan}, {Howell}, {Lord},
  {Marshall}, {Ishida}, {Xu}, {Petric}, {Sanders}, {Surace}, {Appleton},
  {Chan}, {Frayer}, {Inami}, {Khachikian}, {Madore}, {Privon}, {Sturm}, {U}, \&
  {Veilleux}}]{Mazzarella:2012}
{Mazzarella}, J.~M., {Iwasawa}, K., {Vavilkin}, T., {et~al.} 2012, \aj, 144,
  125, \dodoi{10.1088/0004-6256/144/5/125}

\bibitem[{{McAlpine} {et~al.}(2020){McAlpine}, {Harrison}, {Rosario},
  {Alexander}, {Ellison}, {Johansson}, \& {Patton}}]{McAlpine:2020}
{McAlpine}, S., {Harrison}, C.~M., {Rosario}, D.~J., {et~al.} 2020, \mnras,
  494, 5713, \dodoi{10.1093/mnras/staa1123}

\bibitem[{{Mechtley} {et~al.}(2015){Mechtley}, {Jahnke}, {Windhorst}, {Andrae},
  {Cisternas}, {Cohen}, {Hewlett}, {Koekemoer}, {Schramm}, {Schulze},
  {Silverman}, {Villforth}, {van der Wel}, \& {Wisotzki}}]{Mechtley:2015}
{Mechtley}, M., {Jahnke}, K., {Windhorst}, R.~A., {et~al.} 2015, ArXiv
  e-prints.
\newblock \doarXiv{1510.08461}

\bibitem[{{Mihos} \& {Hernquist}(1996)}]{Mihos:1996}
{Mihos}, J.~C., \& {Hernquist}, L. 1996, \apj, 464, 641, \dodoi{10.1086/177353}

\bibitem[{{M{\"u}ller-S{\'a}nchez} {et~al.}(2015){M{\"u}ller-S{\'a}nchez},
  {Comerford}, {Nevin}, {Barrows}, {Cooper}, \&
  {Greene}}]{Mueller-Sanchez:2015}
{M{\"u}ller-S{\'a}nchez}, F., {Comerford}, J.~M., {Nevin}, R., {et~al.} 2015,
  \apj, 813, 103, \dodoi{10.1088/0004-637X/813/2/103}

\bibitem[{{Mundy} {et~al.}(2017){Mundy}, {Conselice}, {Duncan}, {Almaini},
  {H{\"a}u{\ss}ler}, \& {Hartley}}]{Mundy:2017}
{Mundy}, C.~J., {Conselice}, C.~J., {Duncan}, K.~J., {et~al.} 2017, \mnras,
  470, 3507, \dodoi{10.1093/mnras/stx1238}

\bibitem[{{Netzer}(2015)}]{Netzer:2015}
{Netzer}, H. 2015, \araa, 53, 365, \dodoi{10.1146/annurev-astro-082214-122302}

\bibitem[{{Nevin} {et~al.}(2023){Nevin}, {Blecha}, {Comerford}, {Simon},
  {Terrazas}, {Barrows}, \& {V{\'a}zquez-Mata}}]{Nevin:2023}
{Nevin}, R., {Blecha}, L., {Comerford}, J., {et~al.} 2023, \mnras,
  \dodoi{10.1093/mnras/stad911}

\bibitem[{{Noll} {et~al.}(2009){Noll}, {Burgarella}, {Giovannoli}, {Buat},
  {Marcillac}, \& {Mu{\~n}oz-Mateos}}]{Noll:2009}
{Noll}, S., {Burgarella}, D., {Giovannoli}, E., {et~al.} 2009, \aap, 507, 1793,
  \dodoi{10.1051/0004-6361/200912497}

\bibitem[{Pfenniger \& Friedli(1991)}]{Pfenniger:1991}
Pfenniger, D., \& Friedli, D. 1991, Astronomy and Astrophysics, 252, 75

\bibitem[{{Piconcelli} {et~al.}(2010){Piconcelli}, {Vignali}, {Bianchi},
  {Mathur}, {Fiore}, {Guainazzi}, {Lanzuisi}, {Maiolino}, \&
  {Nicastro}}]{Piconcelli2010}
{Piconcelli}, E., {Vignali}, C., {Bianchi}, S., {et~al.} 2010, \apjl, 722,
  L147, \dodoi{10.1088/2041-8205/722/2/L147}

\bibitem[{{Polletta} {et~al.}(2007){Polletta}, {Tajer}, {Maraschi},
  {Trinchieri}, {Lonsdale}, {Chiappetti}, {Andreon}, {Pierre}, {Le F{\`e}vre},
  {Zamorani}, {Maccagni}, {Garcet}, {Surdej}, {Franceschini}, {Alloin},
  {Shupe}, {Surace}, {Fang}, {Rowan-Robinson}, {Smith}, \&
  {Tresse}}]{Polletta:2007}
{Polletta}, M., {Tajer}, M., {Maraschi}, L., {et~al.} 2007, \apj, 663, 81,
  \dodoi{10.1086/518113}

\bibitem[{{Pozzetti} {et~al.}(2010){Pozzetti}, {Bolzonella}, {Zucca},
  {Zamorani}, {Lilly}, {Renzini}, {Moresco}, {Mignoli}, {Cassata}, {Tasca},
  {Lamareille}, {Maier}, {Meneux}, {Halliday}, {Oesch}, {Vergani}, {Caputi},
  {Kova{\v{c}}}, {Cimatti}, {Cucciati}, {Iovino}, {Peng}, {Carollo}, {Contini},
  {Kneib}, {Le F{\'e}vre}, {Mainieri}, {Scodeggio}, {Bardelli}, {Bongiorno},
  {Coppa}, {de la Torre}, {de Ravel}, {Franzetti}, {Garilli}, {Kampczyk},
  {Knobel}, {Le Borgne}, {Le Brun}, {Pell{\`o}}, {Perez Montero},
  {Ricciardelli}, {Silverman}, {Tanaka}, {Tresse}, {Abbas}, {Bottini}, {Cappi},
  {Guzzo}, {Koekemoer}, {Leauthaud}, {Maccagni}, {Marinoni}, {McCracken},
  {Memeo}, {Porciani}, {Scaramella}, {Scarlata}, \& {Scoville}}]{Pozzetti:2010}
{Pozzetti}, L., {Bolzonella}, M., {Zucca}, E., {et~al.} 2010, \aap, 523, A13,
  \dodoi{10.1051/0004-6361/200913020}

\bibitem[{{Qu} {et~al.}(2017){Qu}, {Helly}, {Bower}, {Theuns}, {Crain},
  {Frenk}, {Furlong}, {McAlpine}, {Schaller}, {Schaye}, \& {White}}]{Qu:2017}
{Qu}, Y., {Helly}, J.~C., {Bower}, R.~G., {et~al.} 2017, \mnras, 464, 1659,
  \dodoi{10.1093/mnras/stw2437}

\bibitem[{{Ricci} {et~al.}(2017){Ricci}, {Bauer}, {Treister}, {Schawinski},
  {Privon}, {Blecha}, {Arevalo}, {Armus}, {Harrison}, {Ho}, {Iwasawa},
  {Sanders}, \& {Stern}}]{Ricci:2017}
{Ricci}, C., {Bauer}, F.~E., {Treister}, E., {et~al.} 2017, \mnras,
  \dodoi{10.1093/mnras/stx173}

\bibitem[{{Ricci} {et~al.}(2021){Ricci}, {Privon}, {Pfeifle}, {Armus},
  {Iwasawa}, {Torres-Alb{\`a}}, {Satyapal}, {Bauer}, {Treister}, {Ho}, {Aalto},
  {Ar{\'e}valo}, {Barcos-Mu{\~n}oz}, {Charmandaris}, {Diaz-Santos}, {Evans},
  {Gao}, {Inami}, {Koss}, {Lansbury}, {Linden}, {Medling}, {Sanders}, {Song},
  {Stern}, {U}, {Ueda}, \& {Yamada}}]{Ricci:2021}
{Ricci}, C., {Privon}, G.~C., {Pfeifle}, R.~W., {et~al.} 2021, \mnras, 506,
  5935, \dodoi{10.1093/mnras/stab2052}

\bibitem[{{Rodriguez} {et~al.}(2020){Rodriguez}, {Gonzalez}, {O'Mill},
  {Gazta{\~n}aga}, {Fosalba}, {Lambas}, {Mezcua}, \& {Siudek}}]{Rodriguez:2020}
{Rodriguez}, F., {Gonzalez}, E.~J., {O'Mill}, A.~L., {et~al.} 2020, \aap, 634,
  A123, \dodoi{10.1051/0004-6361/201937215}

\bibitem[{Sakamoto {et~al.}(1999)Sakamoto, Okumura, Ishizuki, \&
  Scoville}]{Sakamoto:1999}
Sakamoto, K., Okumura, S., Ishizuki, S., \& Scoville, N. 1999, The
  Astrophysical Journal, 525, 691

\bibitem[{{Salvato} {et~al.}(2009){Salvato}, {Hasinger}, {Ilbert}, {Zamorani},
  {Brusa}, {Scoville}, {Rau}, {Capak}, {Arnouts}, {Aussel}, {Bolzonella},
  {Buongiorno}, {Cappelluti}, {Caputi}, {Civano}, {Cook}, {Elvis}, {Gilli},
  {Jahnke}, {Kartaltepe}, {Impey}, {Lamareille}, {Le Floc'h}, {Lilly},
  {Mainieri}, {McCarthy}, {McCracken}, {Mignoli}, {Mobasher}, {Murayama},
  {Sasaki}, {Sanders}, {Schiminovich}, {Shioya}, {Shopbell}, {Silverman},
  {Smol{\v{c}}i{\'c}}, {Surace}, {Taniguchi}, {Thompson}, {Trump}, {Urry}, \&
  {Zamojski}}]{Salvato:2009}
{Salvato}, M., {Hasinger}, G., {Ilbert}, O., {et~al.} 2009, \apj, 690, 1250,
  \dodoi{10.1088/0004-637X/690/2/1250}

\bibitem[{{Salvato} {et~al.}(2011){Salvato}, {Ilbert}, {Hasinger}, {Rau},
  {Civano}, {Zamorani}, {Brusa}, {Elvis}, {Vignali}, {Aussel}, {Comastri},
  {Fiore}, {Le Floc'h}, {Mainieri}, {Bardelli}, {Bolzonella}, {Bongiorno},
  {Capak}, {Caputi}, {Cappelluti}, {Carollo}, {Contini}, {Garilli}, {Iovino},
  {Fotopoulou}, {Fruscione}, {Gilli}, {Halliday}, {Kneib}, {Kakazu},
  {Kartaltepe}, {Koekemoer}, {Kovac}, {Ideue}, {Ikeda}, {Impey}, {Le Fevre},
  {Lamareille}, {Lanzuisi}, {Le Borgne}, {Le Brun}, {Lilly}, {Maier},
  {Manohar}, {Masters}, {McCracken}, {Messias}, {Mignoli}, {Mobasher}, {Nagao},
  {Pello}, {Puccetti}, {Perez-Montero}, {Renzini}, {Sargent}, {Sanders},
  {Scodeggio}, {Scoville}, {Shopbell}, {Silvermann}, {Taniguchi}, {Tasca},
  {Tresse}, {Trump}, \& {Zucca}}]{Salvato:2011}
{Salvato}, M., {Ilbert}, O., {Hasinger}, G., {et~al.} 2011, \apj, 742, 61,
  \dodoi{10.1088/0004-637X/742/2/61}

\bibitem[{{Sanders} {et~al.}(1988){Sanders}, {Soifer}, {Elias}, {Madore},
  {Matthews}, {Neugebauer}, \& {Scoville}}]{Sanders:1988}
{Sanders}, D.~B., {Soifer}, B.~T., {Elias}, J.~H., {et~al.} 1988, \apj, 325,
  74, \dodoi{10.1086/165983}

\bibitem[{{Satyapal} {et~al.}(2014){Satyapal}, {Ellison}, {McAlpine}, {Hickox},
  {Patton}, \& {Mendel}}]{Satyapal:2014}
{Satyapal}, S., {Ellison}, S.~L., {McAlpine}, W., {et~al.} 2014, \mnras, 441,
  1297, \dodoi{10.1093/mnras/stu650}

\bibitem[{{Satyapal} {et~al.}(2017){Satyapal}, {Secrest}, {Ricci}, {Ellison},
  {Rothberg}, {Blecha}, {Constantin}, {Gliozzi}, {McNulty}, \&
  {Ferguson}}]{Satyapal:2017}
{Satyapal}, S., {Secrest}, N.~J., {Ricci}, C., {et~al.} 2017, \apj, 848, 126,
  \dodoi{10.3847/1538-4357/aa88ca}

\bibitem[{{Scoville} {et~al.}(2007){Scoville}, {Aussel}, {Brusa}, {Capak},
  {Carollo}, {Elvis}, {Giavalisco}, {Guzzo}, {Hasinger}, {Impey}, {Kneib},
  {LeFevre}, {Lilly}, {Mobasher}, {Renzini}, {Rich}, {Sanders}, {Schinnerer},
  {Schminovich}, {Shopbell}, {Taniguchi}, \& {Tyson}}]{Scoville:2007}
{Scoville}, N., {Aussel}, H., {Brusa}, M., {et~al.} 2007, \apjs, 172, 1,
  \dodoi{10.1086/516585}

\bibitem[{{Secrest} {et~al.}(2015){Secrest}, {Dudik}, {Dorland}, {Zacharias},
  {Makarov}, {Fey}, {Frouard}, \& {Finch}}]{Secrest:2015}
{Secrest}, N.~J., {Dudik}, R.~P., {Dorland}, B.~N., {et~al.} 2015, \apjs, 221,
  12, \dodoi{10.1088/0067-0049/221/1/12}

\bibitem[{Sellwood(1981)}]{Sellwood:1981}
Sellwood, J. 1981, Astronomy and Astrophysics, 99, 362

\bibitem[{{Shah} {et~al.}(2020){Shah}, {Kartaltepe}, {Magagnoli}, {Cox},
  {Wetherell}, {Vanderhoof}, {Calabro}, {Chartab}, {Conselice}, {Croton},
  {Donley}, {de Groot}, {de la Vega}, {Hathi}, {Ilbert}, {Inami}, {Kocevski},
  {Koekemoer}, {Lemaux}, {Mantha}, {Marchesi}, {Martig}, {Masters}, {McGrath},
  {McIntosh}, {Moreno}, {Nayyeri}, {Pampliega}, {Salvato}, {Snyder},
  {Straughn}, {Treister}, \& {Weston}}]{Shah:2020}
{Shah}, E.~A., {Kartaltepe}, J.~S., {Magagnoli}, C.~T., {et~al.} 2020, \apj,
  904, 107, \dodoi{10.3847/1538-4357/abbf59}

\bibitem[{{Silva} {et~al.}(2021){Silva}, {Marchesini}, {Silverman}, {Martis},
  {Iono}, {Espada}, \& {Skelton}}]{Silva:2021}
{Silva}, A., {Marchesini}, D., {Silverman}, J.~D., {et~al.} 2021, \apj, 909,
  124, \dodoi{10.3847/1538-4357/abdbb1}

\bibitem[{{Silverman} {et~al.}(2011){Silverman}, {Kampczyk}, {Jahnke},
  {Andrae}, {Lilly}, {Elvis}, {Civano}, {Mainieri}, {Vignali}, {Zamorani},
  {Nair}, {Le F{\`e}vre}, {de Ravel}, {Bardelli}, {Bongiorno}, {Bolzonella},
  {Cappi}, {Caputi}, {Carollo}, {Contini}, {Coppa}, {Cucciati}, {de la Torre},
  {Franzetti}, {Garilli}, {Halliday}, {Hasinger}, {Iovino}, {Knobel},
  {Koekemoer}, {Kova{\v c}}, {Lamareille}, {Le Borgne}, {Le Brun}, {Maier},
  {Mignoli}, {Pello}, {P{\'e}rez-Montero}, {Ricciardelli}, {Peng}, {Scodeggio},
  {Tanaka}, {Tasca}, {Tresse}, {Vergani}, {Zucca}, {Brusa}, {Cappelluti},
  {Comastri}, {Finoguenov}, {Fu}, {Gilli}, {Hao}, {Ho}, \&
  {Salvato}}]{Silverman:2011}
{Silverman}, J.~D., {Kampczyk}, P., {Jahnke}, K., {et~al.} 2011, \apj, 743, 2,
  \dodoi{10.1088/0004-637X/743/1/2}

\bibitem[{{Simmons} {et~al.}(2012){Simmons}, {Urry}, {Schawinski}, {Cardamone},
  \& {Glikman}}]{Simmons:2012}
{Simmons}, B.~D., {Urry}, C.~M., {Schawinski}, K., {Cardamone}, C., \&
  {Glikman}, E. 2012, \apj, 761, 75, \dodoi{10.1088/0004-637X/761/1/75}

\bibitem[{{Springel} {et~al.}(2005){Springel}, {Di Matteo}, \&
  {Hernquist}}]{Springel:2005a}
{Springel}, V., {Di Matteo}, T., \& {Hernquist}, L. 2005, \apjl, 620, L79,
  \dodoi{10.1086/428772}

\bibitem[{{Steffen} {et~al.}(2022){Steffen}, {Fu}, {Brownstein}, {Comerford},
  {Cruz-Gonz{\'a}lez}, {Dai}, {Drory}, {Gross}, {Negrete}, \&
  {Yan}}]{Steffen:2022}
{Steffen}, J.~L., {Fu}, H., {Brownstein}, J.~R., {et~al.} 2022, arXiv e-prints,
  arXiv:2212.02677.
\newblock \doarXiv{2212.02677}

\bibitem[{{Steinborn} {et~al.}(2016){Steinborn}, {Dolag}, {Comerford},
  {Hirschmann}, {Remus}, \& {Teklu}}]{Steinborn:2016}
{Steinborn}, L.~K., {Dolag}, K., {Comerford}, J.~M., {et~al.} 2016, \mnras,
  458, 1013, \dodoi{10.1093/mnras/stw316}

\bibitem[{{Stemo} {et~al.}(2021){Stemo}, {Comerford}, {Barrows}, {Stern},
  {Assef}, {Griffith}, \& {Schechter}}]{Stemo:2021}
{Stemo}, A., {Comerford}, J.~M., {Barrows}, R.~S., {et~al.} 2021, \apj, 923,
  36, \dodoi{10.3847/1538-4357/ac0bbf}

\bibitem[{{Stern} {et~al.}(2012){Stern}, {Assef}, {Benford}, {Blain}, {Cutri},
  {Dey}, {Eisenhardt}, {Griffith}, {Jarrett}, {Lake}, {Masci}, {Petty},
  {Stanford}, {Tsai}, {Wright}, {Yan}, {Harrison}, \& {Madsen}}]{Stern:2012}
{Stern}, D., {Assef}, R.~J., {Benford}, D.~J., {et~al.} 2012, \apj, 753, 30,
  \dodoi{10.1088/0004-637X/753/1/30}

\bibitem[{{Stickley} \& {Canalizo}(2014)}]{Stickley:2014}
{Stickley}, N.~R., \& {Canalizo}, G. 2014, \apj, 786, 12,
  \dodoi{10.1088/0004-637X/786/1/12}

\bibitem[{{Treister} {et~al.}(2012){Treister}, {Schawinski}, {Urry}, \&
  {Simmons}}]{Treister:2012}
{Treister}, E., {Schawinski}, K., {Urry}, C.~M., \& {Simmons}, B.~D. 2012,
  \apjl, 758, L39, \dodoi{10.1088/2041-8205/758/2/L39}

\bibitem[{{Treister} {et~al.}(2009){Treister}, {Urry}, \&
  {Virani}}]{Treister:2009}
{Treister}, E., {Urry}, C.~M., \& {Virani}, S. 2009, \apj, 696, 110,
  \dodoi{10.1088/0004-637X/696/1/110}

\bibitem[{{Ueda} {et~al.}(2014){Ueda}, {Akiyama}, {Hasinger}, {Miyaji}, \&
  {Watson}}]{Ueda:2014}
{Ueda}, Y., {Akiyama}, M., {Hasinger}, G., {Miyaji}, T., \& {Watson}, M.~G.
  2014, \apj, 786, 104, \dodoi{10.1088/0004-637X/786/2/104}

\bibitem[{{Urrutia} {et~al.}(2008){Urrutia}, {Lacy}, \&
  {Becker}}]{Urrutia:2008}
{Urrutia}, T., {Lacy}, M., \& {Becker}, R.~H. 2008, \apj, 674, 80,
  \dodoi{10.1086/523959}

\bibitem[{van Albada \& Roberts~Jr(1981)}]{VanAldaba:1981}
van Albada, G.~D., \& Roberts~Jr, W. 1981, The Astrophysical Journal, 246, 740

\bibitem[{{Van Wassenhove} {et~al.}(2012){Van Wassenhove}, {Volonteri},
  {Mayer}, {Dotti}, {Bellovary}, \& {Callegari}}]{Van_Wassenhove:2012}
{Van Wassenhove}, S., {Volonteri}, M., {Mayer}, L., {et~al.} 2012, \apjl, 748,
  L7, \dodoi{10.1088/2041-8205/748/1/L7}

\bibitem[{{Villforth} {et~al.}(2014){Villforth}, {Hamann}, {Rosario},
  {Santini}, {McGrath}, {van der Wel}, {Chang}, {Guo}, {Dahlen}, {Bell},
  {Conselice}, {Croton}, {Dekel}, {Faber}, {Grogin}, {Hamilton}, {Hopkins},
  {Juneau}, {Kartaltepe}, {Kocevski}, {Koekemoer}, {Koo}, {Lotz}, {McIntosh},
  {Mozena}, {Somerville}, \& {Wild}}]{Villforth:2014}
{Villforth}, C., {Hamann}, F., {Rosario}, D.~J., {et~al.} 2014, \mnras, 439,
  3342, \dodoi{10.1093/mnras/stu173}

\bibitem[{{Villforth} {et~al.}(2016){Villforth}, {Hamilton}, {Pawlik},
  {Hewlett}, {Rowlands}, {Herbst}, {Shankar}, {Fontana}, {Hamann}, {Koekemoer},
  {Pforr}, {Trump}, \& {Wuyts}}]{Villforth:2016}
{Villforth}, C., {Hamilton}, T., {Pawlik}, M.~M., {et~al.} 2016, ArXiv
  e-prints.
\newblock \doarXiv{1611.06236}

\bibitem[{{Volonteri} {et~al.}(2015){Volonteri}, {Capelo}, {Netzer},
  {Bellovary}, {Dotti}, \& {Governato}}]{Volonteri:2015}
{Volonteri}, M., {Capelo}, P.~R., {Netzer}, H., {et~al.} 2015, \mnras, 452, L6,
  \dodoi{10.1093/mnrasl/slv078}

\bibitem[{{Wang} {et~al.}(2020){Wang}, {Pearson}, \&
  {Rodriguez-Gomez}}]{Wang:2020}
{Wang}, L., {Pearson}, W.~J., \& {Rodriguez-Gomez}, V. 2020, \aap, 644, A87,
  \dodoi{10.1051/0004-6361/202038084}

\bibitem[{{Weigel} {et~al.}(2018){Weigel}, {Schawinski}, {Treister},
  {Trakhtenbrot}, \& {Sanders}}]{Weigel:2018}
{Weigel}, A.~K., {Schawinski}, K., {Treister}, E., {Trakhtenbrot}, B., \&
  {Sanders}, D.~B. 2018, \mnras, 476, 2308, \dodoi{10.1093/mnras/sty383}

\bibitem[{{Weigel} {et~al.}(2017){Weigel}, {Schawinski}, {Caplar}, {Carpineti},
  {Hart}, {Kaviraj}, {Keel}, {Kruk}, {Lintott}, {Nichol}, {Simmons}, \&
  {Smethurst}}]{Weigel:2017}
{Weigel}, A.~K., {Schawinski}, K., {Caplar}, N., {et~al.} 2017, \apj, 845, 145,
  \dodoi{10.3847/1538-4357/aa8097}

\bibitem[{{Weston} {et~al.}(2017){Weston}, {McIntosh}, {Brodwin}, {Mann},
  {Cooper}, {McConnell}, \& {Nielsen}}]{Weston:2017}
{Weston}, M.~E., {McIntosh}, D.~H., {Brodwin}, M., {et~al.} 2017, \mnras, 464,
  3882, \dodoi{10.1093/mnras/stw2620}

\bibitem[{{Wittman} {et~al.}(2016){Wittman}, {Bhaskar}, \&
  {Tobin}}]{Wittman:2016}
{Wittman}, D., {Bhaskar}, R., \& {Tobin}, R. 2016, \mnras, 457, 4005,
  \dodoi{10.1093/mnras/stw261}

\bibitem[{{Wright} {et~al.}(2010){Wright}, {Eisenhardt}, {Mainzer}, {Ressler},
  {Cutri}, {Jarrett}, {Kirkpatrick}, {Padgett}, {McMillan}, {Skrutskie},
  {Stanford}, {Cohen}, {Walker}, {Mather}, {Leisawitz}, {Gautier}, {McLean},
  {Benford}, {Lonsdale}, {Blain}, {Mendez}, {Irace}, {Duval}, {Liu}, {Royer},
  {Heinrichsen}, {Howard}, {Shannon}, {Kendall}, {Walsh}, {Larsen}, {Cardon},
  {Schick}, {Schwalm}, {Abid}, {Fabinsky}, {Naes}, \& {Tsai}}]{Wright:2010}
{Wright}, E.~L., {Eisenhardt}, P.~R.~M., {Mainzer}, A.~K., {et~al.} 2010, \aj,
  140, 1868, \dodoi{10.1088/0004-6256/140/6/1868}

\bibitem[{{Yu} {et~al.}(2011){Yu}, {Lu}, {Mohayaee}, \& {Colin}}]{Yu:2011}
{Yu}, Q., {Lu}, Y., {Mohayaee}, R., \& {Colin}, J. 2011, \apj, 738, 92,
  \dodoi{10.1088/0004-637X/738/1/92}

\end{thebibliography}
\end{document}